\documentclass[a4paper,notitlepage,pra,11pt,aps,showpacs,superscriptaddress]
{revtex4-1}
\usepackage[english]{babel}
\usepackage[utf8]{inputenc}
\usepackage{graphicx,epsfig}

\def\ket|#1>{| #1 \rangle}
\def\bra<#1|{\langle #1 |}
\def\<{\langle}
\def\>{\rangle}
\def\{{\lbrace}
\def\}{\rbrace}
\def\({\left(}
\def\){\right)}
\def\[{\left[}
\def\]{\right]}
\def\sign#1{\mathrm{sign#1}}

\def\beq{\begin{equation}}
\def\eeq{\end{equation}}
\def\ba{\begin{eqnarray}}
\def\ea{\end{eqnarray}}
\def\eqref#1{(\ref{#1})}

\begin{document}

\title{Entanglement over the rainbow}

\author{Giovanni Ram\'{\i}rez}
\affiliation{Instituto de F\'{\i}sica Te\'orica UAM/CSIC, Madrid,
  Spain.}
\email{giovanni.ramirez@uam.es}

\author{Javier Rodr\'{\i}guez-Laguna}
\affiliation{Departamento de F\'{\i}sica Fundamental, UNED, Madrid, Spain.}

\author{Germ\'an Sierra}
\affiliation{Instituto de F\'{\i}sica Te\'orica UAM/CSIC, Madrid,
  Spain.}

\begin{abstract}
  In one dimension the area law for the entanglement entropy is violated
  maximally by the ground state of a strong inhomogeneous spin chain, the so
  called concentric singlet phase (CSP), that looks like a rainbow connecting
  the two halves of the chain. In this paper we show that, in the weak
  inhomogeneity limit, the rainbow state is a thermofield double of a
  conformal field theory with a temperature proportional to the inhomogeneity
  parameter. This result suggests some relation of the CSP with black
  holes. Finally, we propose an extension of the model to higher dimensions.
\end{abstract}

\date{March 11, 2015}
\maketitle
%\tableofcontents

%%%%%%%%%%%%%%%%%%%%%%%%%%%%%%%%%%%%%%%%%%%%%%%%%%%%%%%%%%%%%%%%%%%%%%%%%%%%
\section{Introduction}
\label{sec:introduction}

Entanglement has become a very useful tool to study the structure of complex
quantum states, such as the ground states (GS) of interacting systems
\cite{Amico.RMP.08}. Geometry and quantum structure are linked via the
so-called {\em area laws} \cite{Sredniki.PRL.93, Eisert.RMP.10}, which state
that the entanglement entropy of a block $B$, calculated in the ground state
of a local Hamiltonian, scales with the area of the boundary of the
block. Area laws have been proved only in a few cases, such as 1D gapped
systems \cite{Hastings.JSTAT.07}, and are violated in several interesting
cases, such as conformal systems in 1D \cite{Holzhey.NPB.94,
  Vidal.PRL.03,Calabrese.JSTAT.04}, where the entropy grows logarithmically
with the block size, with a prefactor proportional to the central charge of
the conformal field theory (CFT). These exceptions of the area law have
received an increasing attention in recent studies \cite{Hujse_Swingle.PRB.13,
  Trombettoni.14}.

A strong violation of the area law takes place in an inhomogeneous free
fermion model in 1D where the hopping amplitudes between consecutive sites
decay exponentially outwards from the center of the chain
\cite{Vitagliano.NJP.10}. By tuning the exponential factor, the GS of the
system evolves smoothly from a logarithmic law towards a volume law for the
entanglement entropy between the left and the right halves of the chain
\cite{Ramirez.rainbow.14}. In the strong inhomogeneity regime, when the
exponential factor is high enough, the GS is the product state of Bell-pairs
symmetrically distributed around the center of the system, as shown in
Fig. \ref{fig:RB_scheme}. This valence bond state was termed concentric
singlet phase (CSP) \cite{Vitagliano.NJP.10} or simply, rainbow state
\cite{Ramirez.rainbow.14}. The volume law for strong inhomogeneous chains can
be easily understood from the rainbow picture. The entanglement entropy of
half of the chain is given essentially by the number of bonds connecting the
left and the right halves. However, for weak inhomogeneities the rainbow
picture does not hold because the GS is a resonating valence bond state that
satisfies a volume law plus logarithmic corrections \cite{Ramirez.rainbow.14}.
%
% Smoothed boundary conditions, in which the couplings fall to zero in the
% borders, have been used to reduce the finite-size effects when measuring bulk
% properties of the GS \cite{Vekic.93}. The opposite case, in which the
% couplings increase exponentially or hyperbolically, has also been studied in
% the literature \cite{Okunishi.10, Ueda.09, Ueda.10}.
Another type of inhomogeneity has been used to provide smooth boundary
conditions which improve the convergence of the bulk properties of the GS
\cite{Vekic.93}.  Moreover, an exponential increase of the couplings has been
used in a Kondo-like problem \cite{Okunishi.10} and a hyperbolic increase for
the study of the scaling properties of non-deformed systems \cite{Ueda.09,
  Ueda.10}.

The aim of this paper is to understand the free fermion model in the weak
inhomogeneity regime and its relation to the uniform limit given by a
conformal field theory, namely a massless Dirac fermion with open boundary
conditions. One might think that some scaling limit of the model would
correspond to a perturbation of the underlying CFT.  However, the perturbation
cannot correspond to local operators added to the action, since they do not
give rise to volume law entropies.  Quite surprisingly, the solution of this
puzzle is still provided by CFT: the GS is a sort of thermal state that
satisfies a volume law, with a temperature related to the exponential factor
of the hopping amplitudes. From this perspective, the appearance of a volume
law in the GS of the model is not surprising at all since, after all, it
corresponds to a thermal state. Yet, it comes as a surprise that the state
remains pure. The understanding of this apparent contradiction will bring us
to unexpected territories that we shall start to explore.

The organization of the paper is as follows. After a general reminder of our
model in section \ref{sec:model}, containing an analysis of the strong
inhomogeneity limit, we establish in section \ref{sec:continuum} a continuum
approximation in the vicinity of the homogeneous point, given by a deformation
of the critical Hamiltonian. This deformation affects the single-particle
wavefunctions and the low-energy excitations. In section
\ref{sec:entanglement}, we study the entanglement entropy within the CSP. The
continuum approximation of the previous section allows us to give an
expression for the von Neumann and R\'enyi entropies of the left half of the
system throughout the transition. Moreover, we show how the R\'enyi block
entropies fit to the conformal expressions with varying coefficients. In
section \ref{sec:rainbow2D} we address the question: can the maximal area-law
violation of the deformed hopping Hamiltonian be extended to higher
dimensions? Indeed, we find that a natural extension of the concentric singlet
phase can be found in 2D systems. The article ends with a summary of our
conclusions and ideas for further work. Finally, in appendix \ref{sec:qubism},
{\em qubistic} images \cite{Laguna.NJP.12} are shown to provide useful
information about the entanglement structure.

%%%%%%%%%%%%%%%%%%%%%%%%%%%%%%%%%%%%%%%%%%%%%%%%%%%%%%%%%%%%%%%%%%%%%%%%%
\section{Model and Notation}
\label{sec:model}

\subsection{Strong Disorder Renormalization of the Hopping Model}
\label{sec:SDRG}

Let us consider a strongly inhomogeneous XX-model
\beq
  H_{XX} = \sum_{i=1}^{L-1} K_i (\sigma^x_i\sigma^x_{i+1} +
  \sigma^y_i\sigma^y_{i+1}),
  \label{eq:XX_ham}
\eeq
where the $K_i>0$ are very different in value. We can apply the strong
disorder renormalization group (SDRG) of Ma and Dasgupta \cite{Ma.PRB.80} in
order to obtain the GS. The renormalization prescription is to pick up the
strongest coupling, $K_{max}$, and to establish a singlet bond on top of
it. Then, using second order perturbation theory, one finds the effective
coupling between the two neighbours of the singlet,
\beq
  \tilde K = {K_L K_R \over K_{max}}, 
  \label{eq:RG_XX}
\eeq
where $K_L$ and $K_R$ are, respectively, the left and right couplings to the
maximal one. The renormalization continues by choosing the next largest
coupling and so on. Effective couplings may therefore emerge at long
distances. The success of the SDRG scheme depends on the maximal coupling
$K_{max}$ being always much larger than its neighboring values $K_L$ and
$K_R$.

Using the Jordan-Wigner transformation, we can apply the same tools to study a
fermionic hopping model in 1D with the same features \cite{Ramirez.rxx.14,
  Ramirez.rainbow.14}, reading the couplings $K_i$ as hopping amplitudes,
$J_i$:
\beq
  H = \sum_{i=1}^{L-1} J_i c^\dagger_i c_{i+1} + \mathrm{h.c.}\, ,
  \label{eq:ferm_ham}
\eeq
where $c^\dagger_i$ creates a spinless fermion on site $i$, and is related to
$\sigma^+_i$ via the non-local Jordan-Wigner transformation:
\beq
  c^\dagger_i = \sigma^+_i \prod_{j=1}^{i-1} \sigma^z_j \, . 
  \label{eq:JW} 
\eeq
This non-local transformation points at a modification of the SDRG
prescription to take into account the fermionic nature of the
particles. Effective hoppings between non-contiguous sites are equal to the
corresponding coupling in the XX model multiplied by a sign, $J = (-1)^{n_F}
K$, where $n_F$ is the number of fermions between the two sites
\cite{footnote1}. This rule can be implemented with a simple modification of
the renormalization group (RG) prescription. Since a single fermion is always
added at each RG step,
\beq
  \tilde J = - {J_L J_R \over J_{max}} \, . 
  \label{eq:RG_hop}
\eeq
This implies that the hoppings can be either positive or negative. When they
are positive, a singlet-type {\em bond} is established between both sites, of
the form $\ket|\Psi^->\propto \ket|01>-\ket|10>$. If the hopping is negative,
the corresponding triplet-type {\em anti-bond} is established:
$\ket|\Psi^+>\propto \ket|01>+\ket|10>$. Both types of bonds share many
properties, such as the entanglement. They both represent different flavors of
a Bell pair. The SDRG candidate for the GS of the system is a tensor product
of bond or anti-bond states on the corresponding sites. Indeed, it can be
written as a Fermi state:
\beq
  \ket|GS> = \prod_{k=1}^{L/2} d^\dagger_k \ket|0> , 
  \label{eq:Slater}
\eeq 
where $d^\dagger_k$ creates either a bond or an anti-bond on a pair of sites,
i.e.: $d^\dagger_k \propto c^\dagger_i \pm c^\dagger_j$.

It was proved in \cite{Ramirez.rxx.14}, that the Hamiltonian
\eqref{eq:ferm_ham} has the following properties, for any values of the
$\{J_i\}$. It presents particle-hole symmetry, which implies that for every
single-particle eigenstate with energy $\epsilon$ there is another eigenstate
with energy $-\epsilon$, which is related by swapping the sign of the
components of all odd sites. Thus, the ground state takes place at
half-filling. Moreover, the occupation number of every site is equal:
$\<n_i\>=1/2$. This is a non-intuitive result, given the inhomogeneity of
Hamiltonian \eqref{eq:XX_ham}, and it does not hold for excited states.

\subsection{The Rainbow State}
\label{sec:CSP}

Let us describe the family of local Hamiltonians whose ground state approaches
asymptotically the concentric singlet phase, also known as rainbow state, and
give some heuristic arguments to explain the volume law for the entanglement
entropy.

\begin{figure}[h]
  \centering
  \includegraphics[width=120mm]{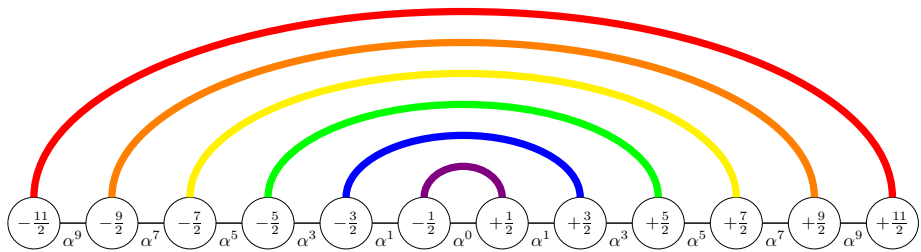}
  \caption{Rainbow state showing the $(-k,+k)$ bonds above the central
    link. Thus, the entanglement entropy between the left and the right halves
    of the chain is $L \log 2$.}
  \label{fig:RB_scheme}
\end{figure}

Let us consider a chain with $2L$ sites, which we will label using half-odd
integers, $n= \pm \frac{1}{2}, \pm \frac{3}{2}, \cdots, \pm \(L
-\frac{1}{2}\)$. On each site $n$, let $c_n$ and $c^\dagger_n$ denote the
annihilation and creation operators of a spinless fermion. The hopping
Hamiltonian is given by \cite{footnote2}
\beq
  H  \equiv  -\frac{J_0}{2} c_{\frac{1}{2}}^\dagger c_{-\frac{1}{2}}
  -\sum_{n=\frac{1}{2}}^{L -\frac{3}{2}} \frac{J_n}{2} \[ c^\dagger_n c_{n+1}
  +c^\dagger_{-n} c_{-(n+1)} \] +\mathrm{h.c.}\,,
  \label{eq:RB_H}
\eeq
where $J_n$ are the hopping amplitudes parametrized as (see
Fig. \ref{fig:RB_scheme} for an illustration)

\beq
  \left\{ 
    \begin{array}{lcrl} 
      J_0(\alpha) &=& 1,  &  \\ 
      J_n(\alpha) &=& \alpha^{2 n}, & n= \frac{1}{2},  \dots,  L-
      \frac{3}{2}.
    \end{array}
  \right.
  \label{eq:RB_couplings} 
\eeq

For $\alpha=1$, this is the uniform 1D spinless fermion model with open
boundary conditions (OBC). The model with $0<\alpha<1$ was introduced by
Vitagliano and coworkers to illustrate a violation of the area law for local
Hamiltonians \cite{Vitagliano.NJP.10} and was studied in much more detail in
reference \cite{Ramirez.rainbow.14}.

In the $\alpha\to 0^+$ limit, the couplings become strongly inhomogeneous and,
as argued in \cite{Ramirez.rainbow.14}, we can employ the SDRG described in
the previous section. The largest hopping is the central one, $J_0=1$, and
gets renormalized to $\tilde J_0 = -\alpha^2$, larger (in absolute value) than
$\alpha^3$, which comes next. The values of $J_n$ are engineered so that this
situation repeats itself for all steps of SDRG, so the bonds are established
in a concentric way around the center, joining sites $+k$ and $-k$, and giving
rise to the aforementioned CSP, as in Fig. \ref{fig:RB_scheme}. Moreover,
notice that the signs alternate.

It is worth to notice the striking similarity between our system and the Kondo
chain \cite{Wilson.75}. Indeed, let us divide our inhomogeneous chain into
three parts: central link, left sites and right sites. The left and right
sites correspond, in our analogy, to the spin up and down chains used in
Wilson's chain representation of the Kondo problem. In both cases, they form a
system of free fermions, with exponentially decaying couplings. In the Kondo
chain, notwithstanding, the central link becomes a magnetic impurity, which
renders the full system non-gaussian.

\subsection{Strong Inhomogeneity Limit}
\label{sec:StrongLimit}

The $\alpha\to 0^+$ limit of \eqref{eq:RB_H}, leading to the rainbow state, is
singular: the Hamiltonian decouples in that limit, and only the central link
survives. Let us consider a very small but non-zero $\alpha$, and study the GS
to first order in perturbation theory. The Hamiltonian is always free, so the
GS is a Slater determinant \eqref{eq:Slater}. The orbital operators,
$d^\dagger_k$, can be expanded in terms of the local creation operators:

\beq
  d^\dagger_k = \sum_i \psi^k_i c^\dagger_i ,
  \label{eq:orbitals}
\eeq
where $\psi^k_i$ are the wavefunction components for the single-body
associated problem, i.e., eigenvectors of the hopping matrix. Let us propose
the following form for them:
\beq
  \label{eq:structure}
  \matrix{
    & & \cdots & -\frac{5}{2} & -\frac{3}{2} & -\frac{1}{2} &
    +\frac{1}{2} & +\frac{3}{2} & +\frac{5}{2} & \cdots \cr
    \psi^1 &=& \cdots & \cdots & \alpha & 1 & 1 & \alpha & \cdots & \cdots \cr
    \psi^2 &=& \cdots & \alpha & 1 & \alpha & -\alpha & -1 & -\alpha & \cdots \cr
    \psi^3 &=& \alpha & 1 & \alpha & 0 & 0 & \alpha & 1 & \alpha \cr
  }
\eeq

Notice the sign alternation, due to the negative sign in the renormalization
prescription \eqref{eq:RG_hop}. It is straightforward to check that all those
$\psi^k$ are eigenstates of the hopping matrix to first order in $\alpha$. We
can now define two families of states: the bonding and the anti-bonding
creating operators, defined as:
\ba
  \nonumber
  \(b^+_{ij}\)^\dagger&=&{1\over\sqrt{2}} \( c^\dagger_i + c^\dagger_j \),
  \\\label{eq:bonding}
  \(b^-_{ij}\)^\dagger&=&{1\over\sqrt{2}} \( c^\dagger_i - c^\dagger_j \).
\ea

In the limit $\alpha\to 0^+$, the GS of Hamiltonian \eqref{eq:RB_H} can be
written as the concentric singlet state or rainbow state:
\beq
  \label{eq:rainbow}
  \ket|R_L>  \equiv  \(b^{s_L}_{-L+\frac{1}{2},L-\frac{1}{2}}\)^\dagger
  \cdots \(b^+_{-\frac{5}{2},\frac{5}{2}}\)^\dagger
  \(b^-_{-\frac{3}{2},\frac{3}{2}}\)^\dagger
  \(b^+_{-\frac{1}{2},\frac{1}{2}}\)^\dagger \ket|0> .
\eeq
where $s_L=(-1)^L$.

%%%%%%%%%%%%%%%%%%%%%%%%%%%%%%%%%%%%%%%%%%%%%%%%%%%%%%%%%%%%%%%%%%%%%%%%%%%%%

\section{Weak Inhomogeneity Limit: Continuum Approximation}
\label{sec:continuum}

The study of the weak inhomogeneity limit, $\alpha\to 1^-$, motivates the
derivation of a continuum approximation of the Hamiltonian
\eqref{eq:RB_H}. This is obtained by expanding the local operator $c_n$ into
the slow modes, $\psi_R(x)$ and $\psi_L(x)$ around the Fermi points $\pm k_F$
\beq
  \frac{c_n}{\sqrt{a}} \simeq e^{i k_F x} \psi_L(x) +e^{-i k_F x} \psi_R(x),
  \label{eq:ContinuumLimit}
\eeq 
located at the position $x=an \in (-{\cal L}, {\cal L})$, where $a$ is the
lattice spacing and ${\cal L} = a L$. In the continuum limit, $a \rightarrow
0$ and $L \rightarrow \infty$, with ${\cal L}$ kept constant. At half-filling,
$k_F=\pi/(2a)$ is the Fermi momentum.
 
Equation \eqref{eq:ContinuumLimit} is the familiar expansion used in the
uniform case, $\alpha=1$, that will enable us to derive the numerical results
found in \cite{Ramirez.rainbow.14}. Plugging Eq. \eqref{eq:ContinuumLimit}
into Eq. \eqref{eq:RB_H} one obtains
\beq
  H \simeq  \frac{ia}{2} \int_{-\cal L}^{\cal L} dx \,
  e^{-\frac{h|x|}{a}} 
  \[ \psi^\dagger_R \partial_x \psi_R -(\partial_x
  \psi^\dagger_R) \psi_R -\psi^\dagger_L \partial_x \psi_L \right.
  +\left. (\partial_x \psi^\dagger_L) \psi_L \],
\label{eq:H_approx}
\eeq
where
\beq
  \alpha = e^{-h/2}.  
  \label{DefAlpha}
\eeq

To derive equation \eqref{eq:H_approx}, we have assumed that the fields
$\psi_{R,L}(x)$ vary slowly with $x$, so that cross terms like $(-1)^{x/a}
\psi_R^\dagger(x) \psi_L(x)$ can be dropped. We have also made a gradient
expansion $\psi(x+a) \simeq \psi(x) + a \partial_x \psi(x)$, keeping only
terms up to the first derivative. The Hamiltonian \eqref{eq:H_approx}
describes the low energy excitations of the original lattice Hamiltonian at
half-filling. It is worth to mention that \eqref{eq:H_approx} is a Hermitian
operator, that is, $H^\dagger = H$, which is of course a consequence of the
hermiticity of \eqref{eq:RB_H}.  In the continuum limit we shall take $h
\rightarrow 0$, with $h/a$ kept constant, so that $\alpha \rightarrow 1^-$.

The boundary conditions (BC) satisfied by the fields $\psi_{L,R}(x)$ at $x =
\pm {\cal L}$, can be derived from equation \eqref{eq:ContinuumLimit} setting
$c_{\pm (L+ \frac{1}{2})} =0$ and taking a continuum limit that yields
\beq
  \label{eq:BoundaryCond}
  \psi_R (\pm {\cal L}) = \mp i \,  \psi_L (\pm {\cal L}).
\eeq
Then, integrating by parts, one can write Hamiltonian \eqref{eq:H_approx} as
\beq
  H \simeq  i a \int_{-\cal L}^{\cal L} dx \,  e^{-\frac{h |x|}{a}} \[
  \psi^\dagger_R \partial_x \psi_R -\psi^\dagger_L \partial_x \psi_L
  \right. 
  -\left. \frac{h}{2 a} \sign{(x)} (\psi^\dagger_R \psi_R
    -\psi^\dagger_L \psi_L ) \]. 
\label{eq:H_integrated} 
\eeq
The Fermi velocity is then given by $v_F= a$, that we set equal to one by
convention (similarly, we replace $h/a \rightarrow h$). The single-body
spectrum of the uniform model, that is, $h=0$, can be easily found
\beq
  E_m = \frac{\pi(m+ 1/2)}{2 L}, \quad m=0, \pm 1, \ldots
  \label{eq:SBspectra}
\eeq 
For the non-uniform model we have the Eqs. 
\beq
  i e^{-h |x|} \[ \partial_x  \mp \frac{h}{2} \, \sign{(x)} \] \psi_{R, L}(x)
  \quad =\quad \pm E  \psi_{R,L}(x),
  \label{eq:EigenValEQ}
\eeq
whose solution is
\beq
  \psi_{R, L}(x) = A_{R,L} e^{h |x|/2} \exp{\[ \mp \frac{iE}{h} \sign{(x)} \(
    e^{h |x|} -1 \) \]}.
  \label{eq:EigenFunc}
\eeq

Notice that, in the limit $h \to 0$, one recovers the usual plane-wave
solutions $\psi_{R,L} \rightarrow A_{R,L} e^{ \mp i E x}$. The BC
\eqref{eq:BoundaryCond} imply:
\ba 
  \nonumber
  A_R\, \exp{\[ -\frac{iE}{h} \( e^{hL} -1 \) \]} &=& -i\, A_L\, \exp{\[
    \frac{iE}{h} \( e^{hL} -1 \) \]}, \\\label{eq:CoeffRelation}
  A_R\, \exp{\[ \frac{iE}{h} \( e^{hL} -1 \) \]} &=& \quad i\, A_L\, \exp{\[
    -\frac{iE}{h} \( e^{hL} -1 \) \]}.
\ea
which, eliminating $A_{R,L}$, yields
\beq
  \label{eq:CoeffReduced}
  \exp{\[ \frac{4iE}{h} \( e^{hL} -1 \) \]} = -1.
\eeq

\begin{figure}[h]
  \centering
  \includegraphics[width=90mm]{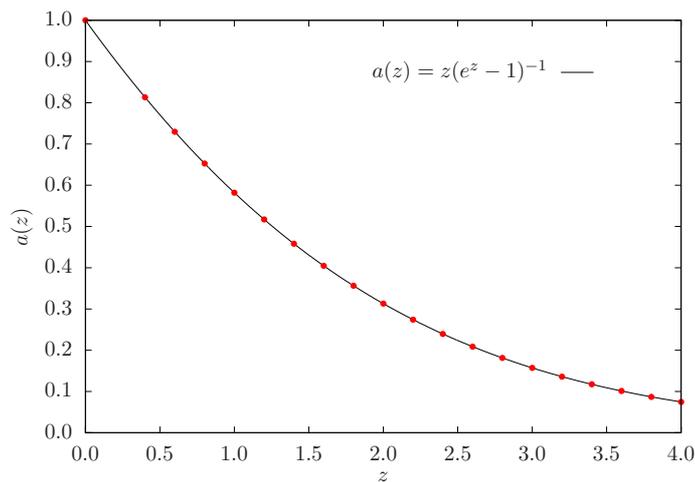}
  \caption{The scaling function $a(z)$ that gives the spectrum of the
    Hamiltonian \eqref{eq:RB_H} near the Fermi energy. The points are the
    numerical data, i.e., the slope of the spectrum at the Fermi point, and
    the continuous line is the analytic result.}
  \label{fig:VFermi}
\end{figure}

The eigenmodes are then given by
\beq
  \label{eq:EigenModes}
  E_m = \frac{h\pi (m+1/2)}{2(e^{hL} -1)} = a(z) \frac{\pi(m+ 1/2)}{2 L},
  \quad m=0, \pm 1, \dots 
\eeq
where $a(z)$, defined as 
\beq
  \label{eq:Velocity}
  a(z) = \frac{z}{e^z -1}, \qquad z \equiv h L.
\eeq
was interpreted in reference \cite{Ramirez.rainbow.14} as the Fermi
velocity. Notice that $z$ has a finite value in the continuum limit since it
can be written as $z = (h/a) {\cal L}$. In the latter reference it was shown
that the single-particle spectrum of the Hamiltonian \eqref{eq:RB_H}
corresponds to Eq. \eqref{eq:EigenModes} with a function $a(z)$ given
numerically in Fig. \ref{fig:VFermi}. As we can see, the analytic expression
\eqref{eq:Velocity} \cite{footnote3} gives a very good fit of the numerical
data.

\begin{figure}
  \centering
  \begin{minipage}[t]{60mm}
    \includegraphics[width=60mm]{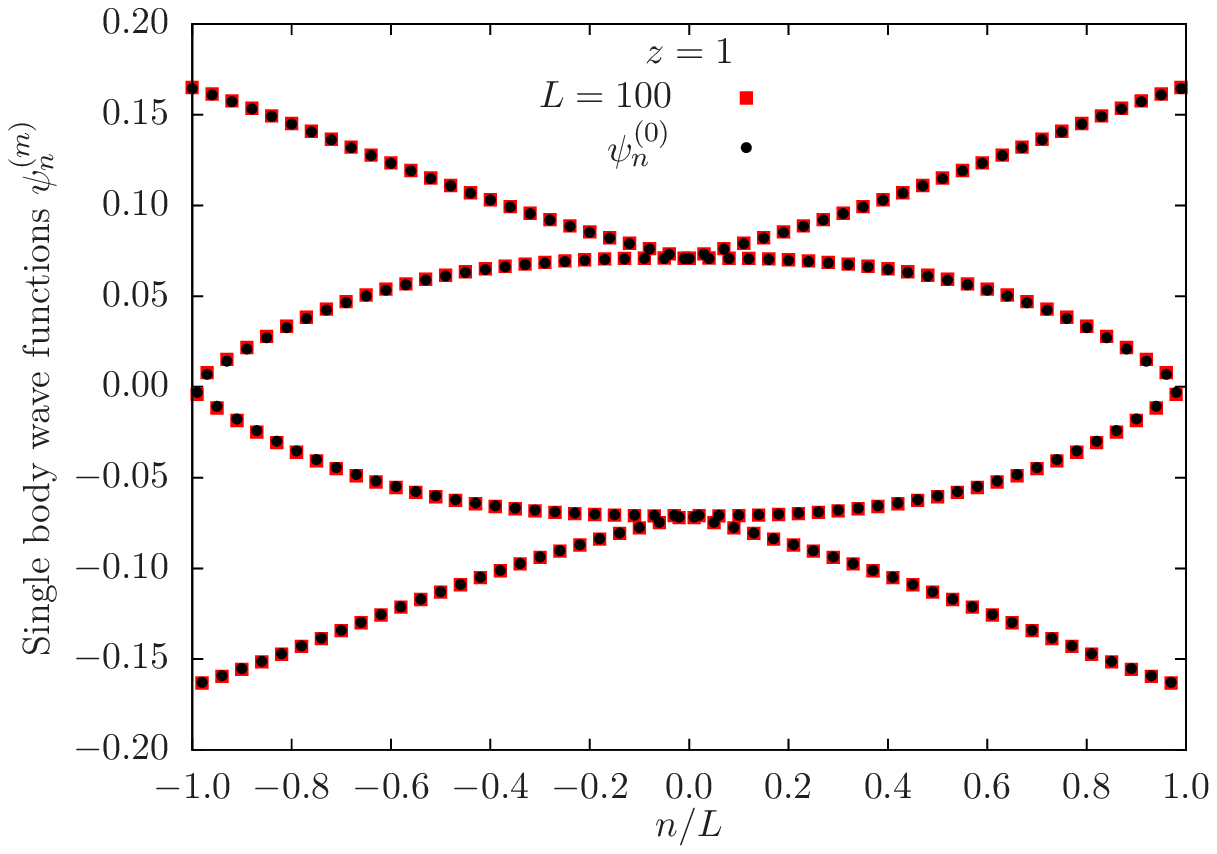} 
  \end{minipage}\hspace{5mm}
  \begin{minipage}[t]{60mm}
    \includegraphics[width=60mm]{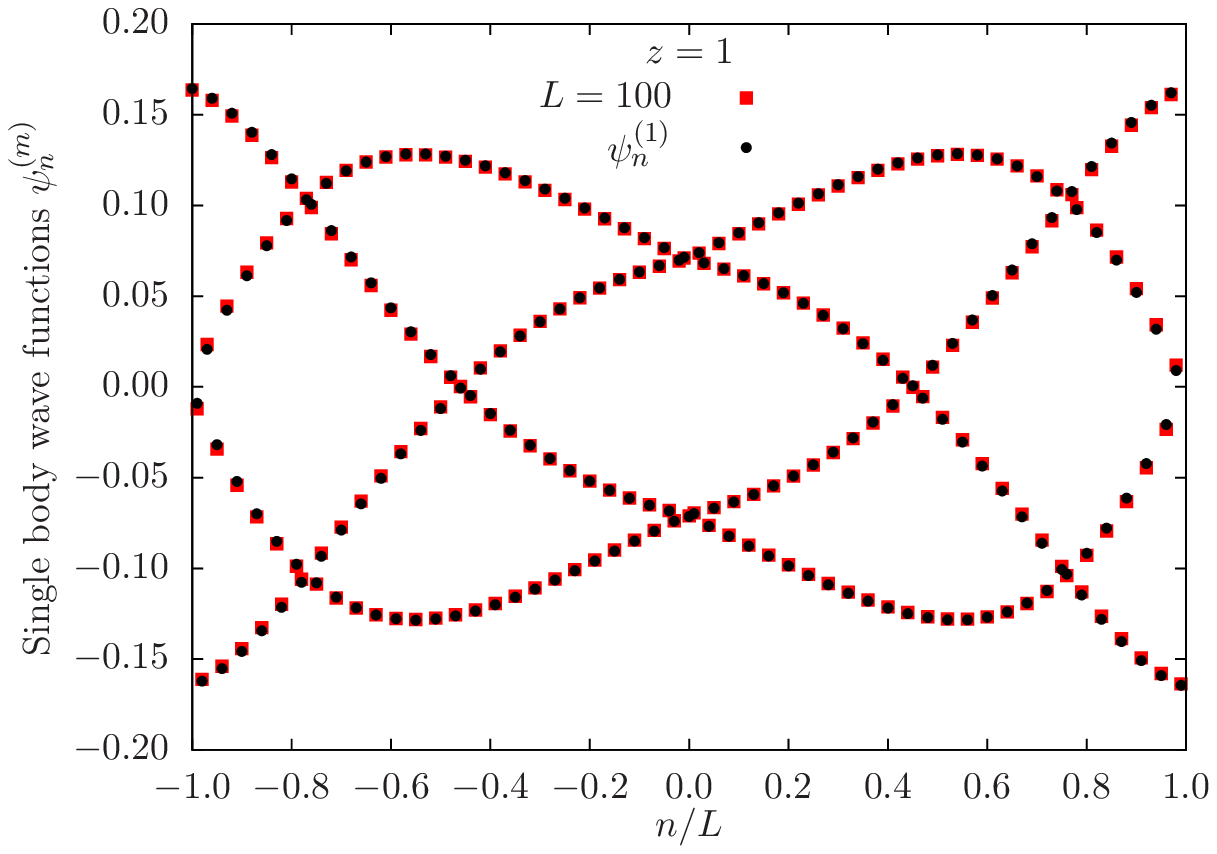}
  \end{minipage}\\\vspace{3mm}
  \begin{minipage}[t]{60mm}
    \includegraphics[width=60mm]{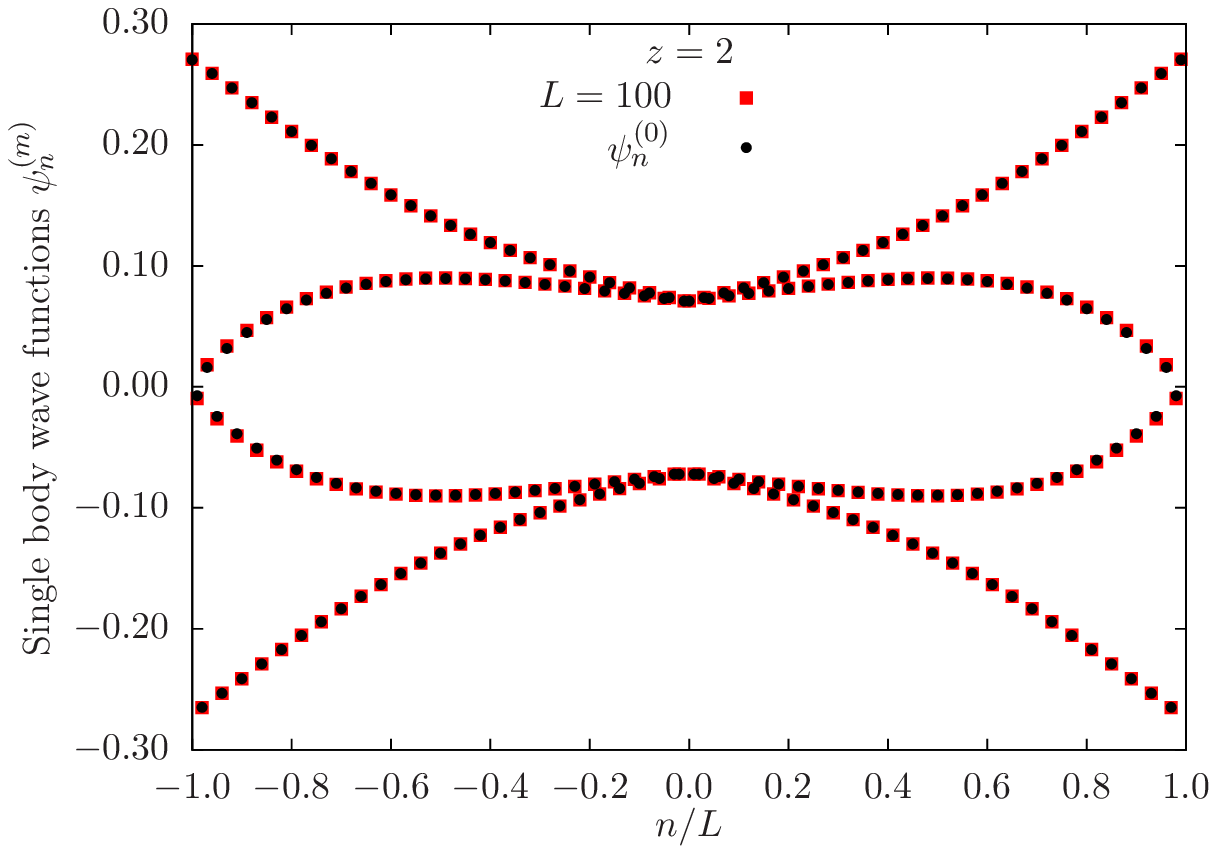}
  \end{minipage}\hspace{5mm}
  \begin{minipage}[t]{60mm}
    \includegraphics[width=60mm]{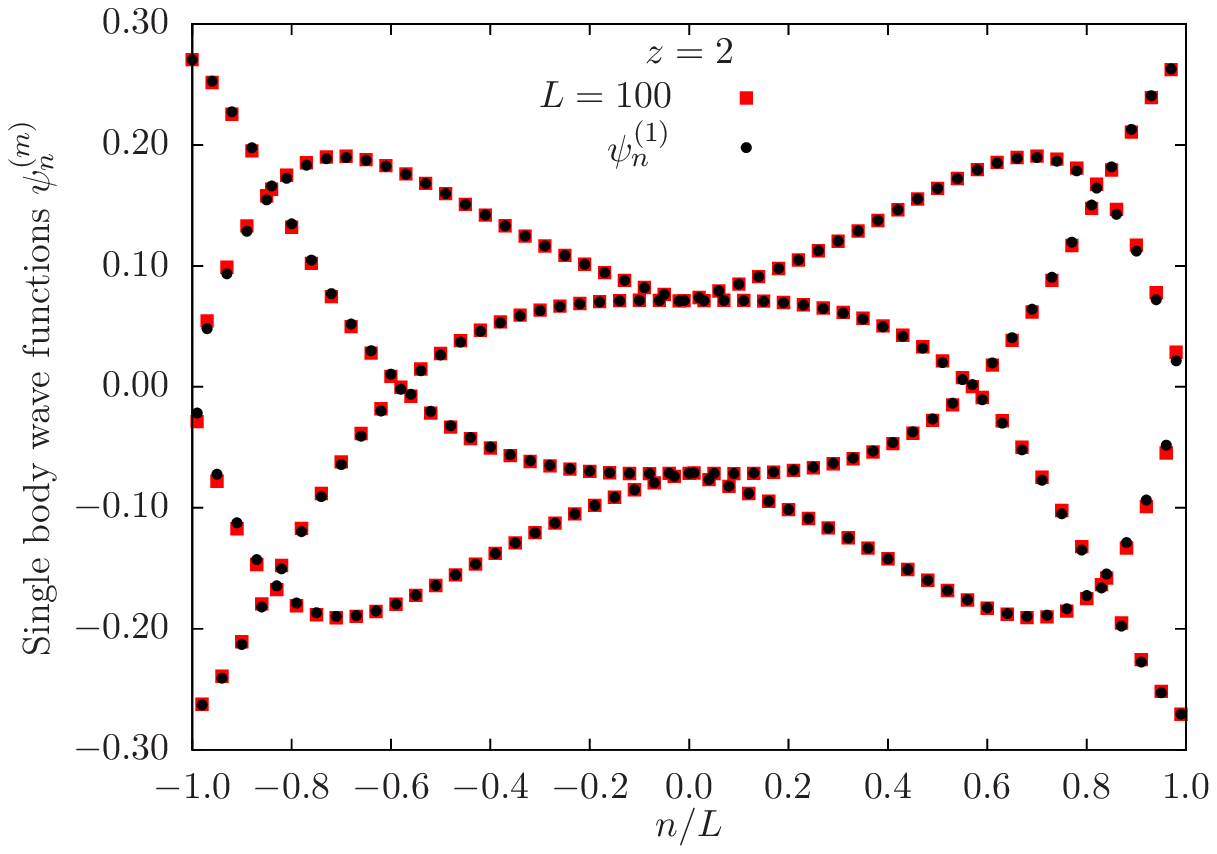}
  \end{minipage}
  \caption{Wavefunctions just below the Fermi level ($m=0$, left column) and
    the next below the Fermi level ($m=1$, right column) for $z=1$ (top) and
    $2$ (bottom). The analytic values of $\psi^{(m)}_n$ are given by
    Eq. \eqref{eq:SingleBodyWF}.}
  \label{fig:WFz}
\end{figure}

To find the eigenfunctions with energy $E_m$, we first compute the constants
$A_{R,L}$ using \eqref{eq:CoeffRelation}
\beq
  A_{R,L} = 
  \exp \left\{\pm i \[ \frac{E_m}{h} \( e^{hL} -1 \) -\frac{\pi}{4} \]\right\},
  \label{eq:Coefs}
\eeq 
and Eq. \eqref{eq:ContinuumLimit}, obtaining
\beq
  \label{eq:SingleBodyWF}
  \psi_n^{(m)} \simeq e^{h |n|/2} \cos \[ \frac{\pi (n-m)}{2}  +\sign{(n)}
  \frac{\pi (m +1/2) }{2} \frac{e^{h |n|} -1} {e^{h L} -1} \],
\eeq
where $n=\pm\frac{1}{2},\cdots,\pm\(L-\frac{1}{2}\)$ and $m=0,\pm
1,\cdots$. Fig. \ref{fig:WFz} shows the numerical and analytic values of
$\psi_n^{(m)}$ for $m=0,1$ and $z=1$ and $2$. As $hn=zn/L$, we see that for
the same value of $z$, all the curves collapse when expressed in the scaled
variable $n/L$.

The results obtained so far suggest that the continuum Hamiltonian
\eqref{eq:H_integrated} can be brought to the standard canonical form of a
free fermion with OBC. To show that this is indeed the case, let us make the
change of variables
\beq
  \tilde x = \sign{(x)} \frac{e^{ h |x|} -1}{h},
  \label{eq:Transfx}
\eeq
that maps the interval $x \in [-L, L]$ into the interval $\tilde{x} \in [-
\tilde{L}, \tilde{L}]$ where
\beq
  \tilde{L} = \frac{e^{h L} -1}{h}. 
  \label{eq:TransfL}
\eeq

The fermion fields in the variable $\tilde x$ are given by
\beq
  \tilde{\psi}_{R,L} (\tilde{x})  =  \( \frac{d\tilde{x} }{d x} \)^{-1/2}
  \psi_{R,L}(x) = e^{-h |x|/2} \psi_{R,L}(x),
  \label{eq:TransfWF}
\eeq
that plugged into \eqref{eq:H_integrated} gives (recall that we set $a=1$, so
${\cal L} = L$)
\beq
  H \simeq i \int_{- \tilde L}^{\tilde L} d \tilde{x} \, \[
  \tilde{\psi}^\dagger_R \partial_{\tilde x} \tilde{\psi}_R
  -\tilde{\psi}^\dagger_L \partial_{\tilde{x}} \tilde{\psi}_L \].
  \label{eq:TransfH}
\eeq 
That is just the free fermion Hamiltonian for a chain of length $2
\tilde{L}$. This result suggests that one could try to derive some of the
properties of the rainbow Hamiltonian, Eq. \eqref{eq:RB_H}, from those of the
free fermion system. This will be done in the next section when discussing the
entanglement properties of the GS.

Notice that Eq. \eqref{eq:Transfx} is not analytic at $x=0$, but if we take $x
> 0$ we obtain
\beq
  \tilde x={e^{hx}-1\over h},
  \label{eq:Transfx_conformal}
\eeq
which is a conformal transformation (similarly, $\tilde x=-(e^{-hx}-1)/h$ if
$x<0$). If we add the euclidean time coordinate, that is, $x\to x+i\tau$, the
transformation \eqref{eq:Transfx_conformal} becomes periodic in $\tau$ with a
period equal to $\beta = 2 \pi/h$. This result leads us to associate to the
system an effective temperature
\beq
  T=\frac{h}{2\pi}.
  \label{eq:effective_temp}
\eeq
This result will be interpreted below.

\subsection{Validity of the continuum approximation}
\label{sec:Validity}

As can be seen in Figs. \ref{fig:VFermi} and \ref{fig:WFz}, the continuum
approximation provides very accurate predictions regarding the wavefunctions
near the Fermi point and the Fermi velocity. Indeed, that is the expected
range of validity of any continuum limit: the long-distance physics which
takes place near the Fermi point.

\begin{figure}
  \epsfig{file=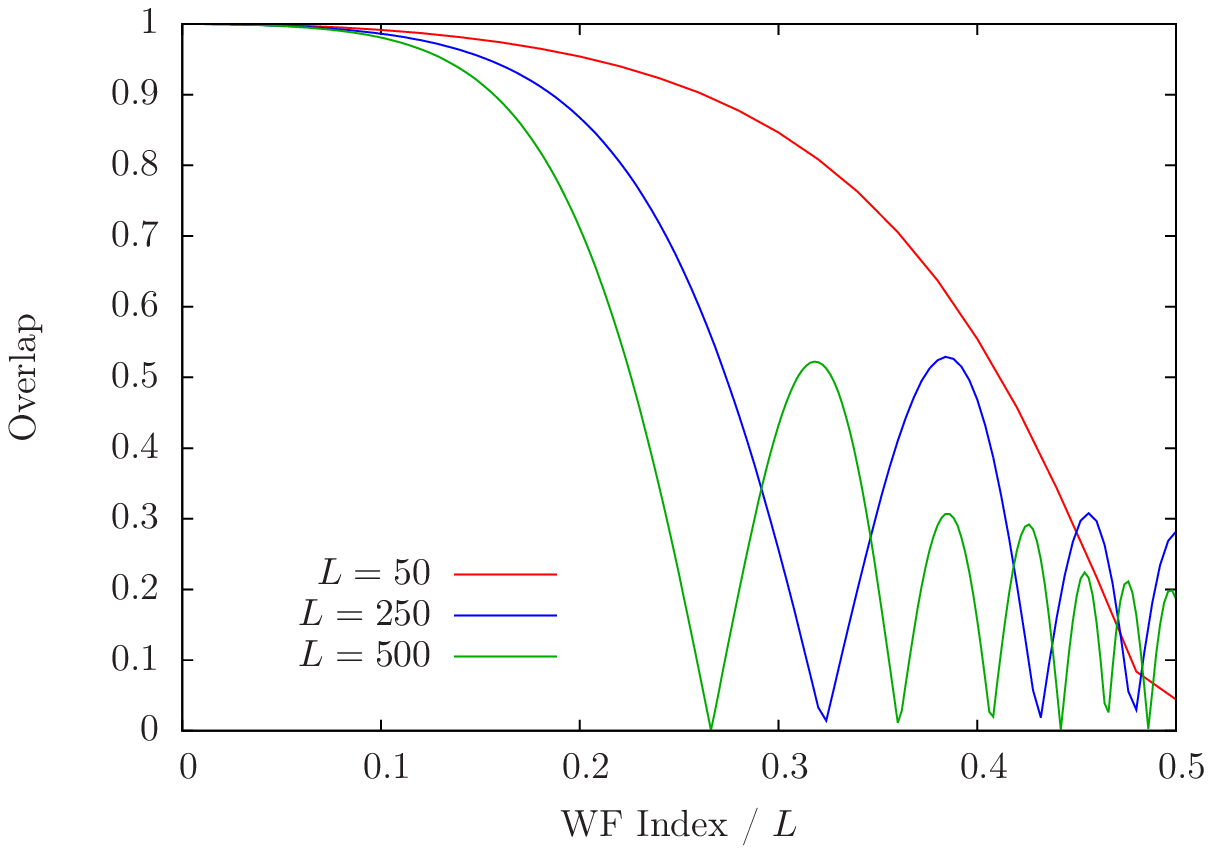,width=7cm}
  \epsfig{file=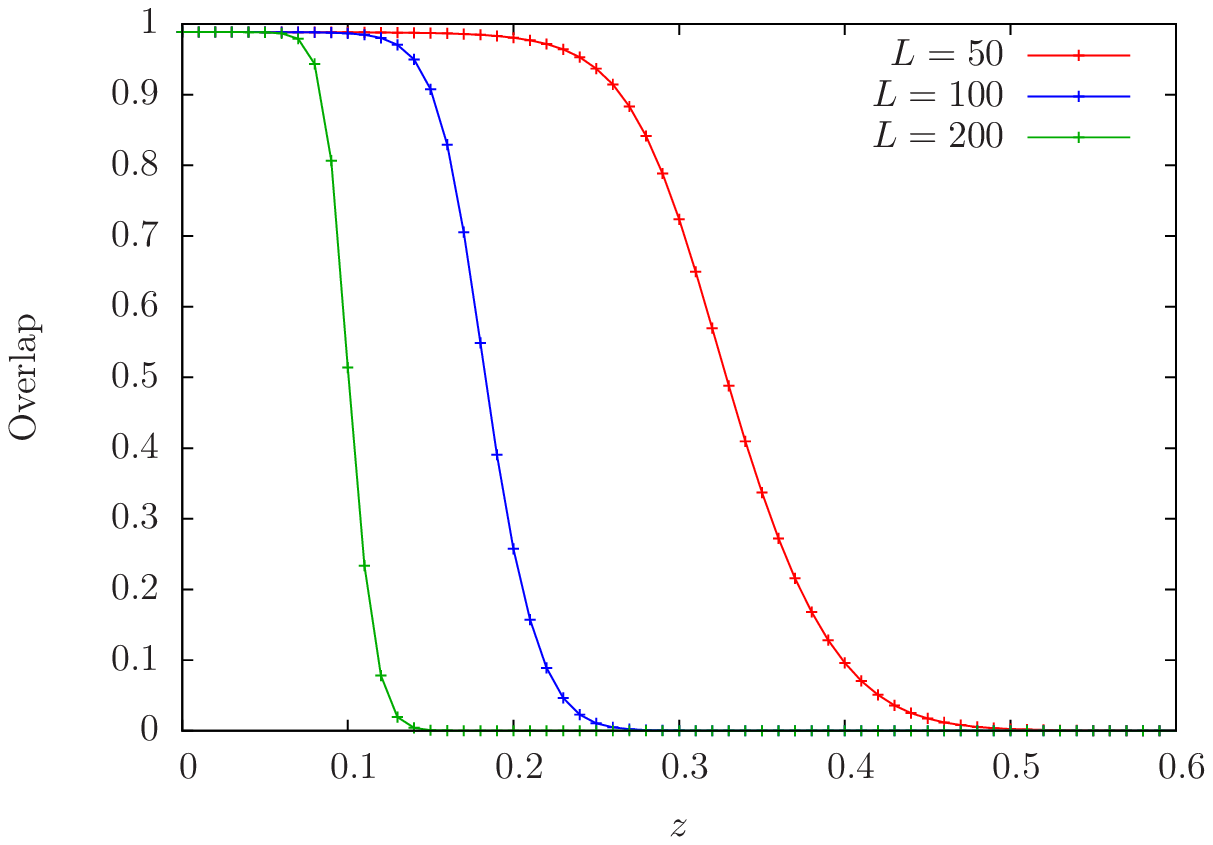,width=7cm}
  \epsfig{file=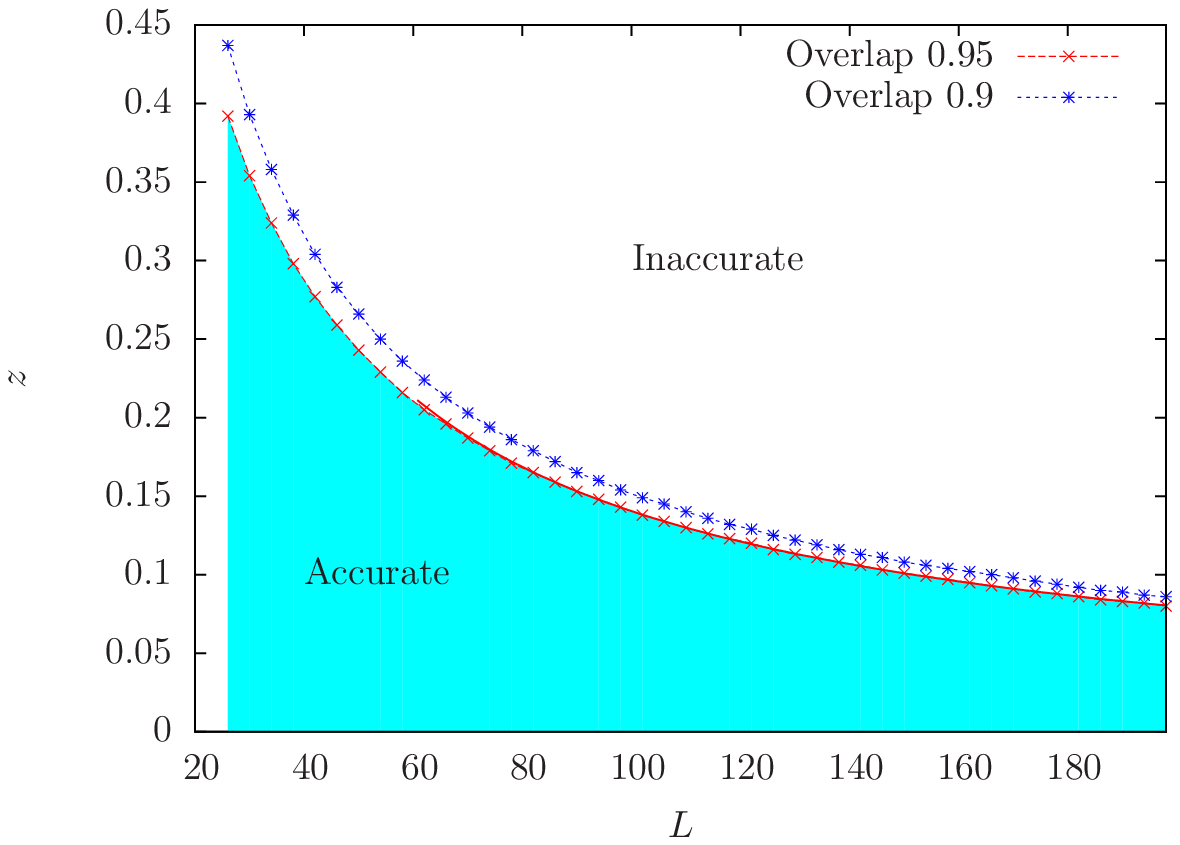,width=8cm}
  \caption{Top left: overlap between continuum approximation and numerical
    single-particle wavefunctions, in a one-to-one basis, for $z=1$. Top
    right: overlap between continuum and numerical many-body states, as a
    function of $z$, for different values of $L$. Bottom: validity region for
    the continuum approximation, showing the lines of $90\%$ and $95\%$
    overlap between the continuum and the exact wavefunctions.}
  \label{fig:cont_error}
\end{figure}

We have explored the limits of the validity of the continuum
approximation. Fig. \ref{fig:cont_error} shows the overlap between the
predicted and the numerical single-particle wavefunctions as we go deeper
beneath the Fermi surface. The horizontal axis shows the rescaled wavefunction
index $m/L$, which is $0$ for the Fermi level and $1$ for the deepest one. The
vertical axis corresponds to the overlap between the continuum approximation
and the actual wavefunction, defined as
\beq 
  {\cal O} = \left| \bra<\psi_{cont}|\left.\psi_{exact}\right\rangle \right|.
  \label{eq:def_overlap}
\eeq

The numerical experiments were performed for $L$ in the range of $50$ to $500$
and $z=1$. Notice that for small wavefunction index $m/L$, the overlap is
virtually one, but it decreases very fast behind a certain critical value. The
explanation is that single-body wavefunctions which are deep below the Fermi
energy vary over very small length scales, rendering the continuum
approximation inaccurate.

Even if the wavefunctions are not correctly predicted in a one-to-one basis,
the complete Slater determinant composing the states can be similar. This
possibility is checked in the top-right panel of Fig. \ref{fig:cont_error},
where we plot the overlap between the full Slater determinant states
(continuum limit and numerical computation) as a function of $z$ for different
values of the system size $L$. We can see that below a certain critical $z$,
the overlap stays close to one, and then it decreases to zero.

The bottom panel of Fig. \ref{fig:cont_error} shows the region of validity of
the continuum approximation in the $(L,z)$ plane, by depicting two lines which
mark the level 0.95 and the level 0.9 for the overlap between the numerical GS
of the Hamiltonian and the continuum approximation obtained by the deformed
uniform wavefunctions.

%%%%%%%%%%%%%%%%%%%%%%%%%%%%%%%%%%%%%%%%%%%%%%%%%%%%%%%%%%%%%%%%%%%%%%%%%%%%

\section{Entanglement Structure}
\label{sec:entanglement}

The most relevant fact about the entanglement structure of the GS of
Eq. \eqref{eq:RB_H} is that the von Neumann entropy scales linearly with the
system size --i.e., with a volume law-- as soon as $\alpha<1$, with a
prefactor that was determined numerically in \cite{Ramirez.rainbow.14} as
$\approx -\log(\alpha)/6$. Can this prefactor be explained?

\subsection{Entanglement in the continuum approximation}
\label{sec:entanglement_continuum}

\begin{figure}
  \centering
  \epsfig{file=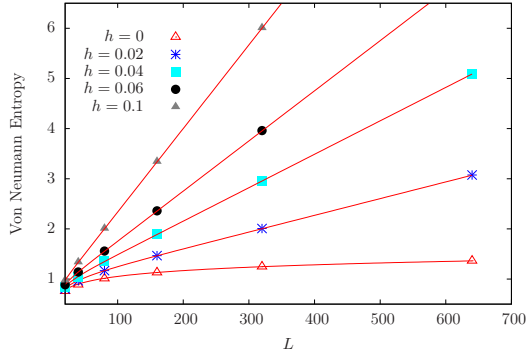,width=8cm}
  \caption{Comparing the half-chain entropy for different $L$ and $h$ with the
    theoretical prediction, Eq. \eqref{eq:S_prediction}.}
  \label{fig:S_prediction}
\end{figure}

The von Neumann entropy of the left half of a critical system with open
boundary conditions and size $L$ is given by
\beq
  S_{CFT}(L) = {c\over 6} \log(L) + c',
  \label{eq:S_conformal}
\eeq
where $c$ is the central charge of the model, and $c'$ is an additive constant
that includes the boundary entropy plus non-universal contributions
\cite{Vidal.PRL.03,Calabrese.JSTAT.04}. For the free fermionic system under
study, $\alpha=1$, we have $c=1$. Taking in combination
Eqs. \eqref{eq:S_conformal} and \eqref{eq:TransfL}, we can provide a
prediction for the entropy of the half-chain in the deformed GS of
\eqref{eq:RB_H}. Indeed, substituting $L$ by $\tilde L$, we obtain
\beq
  S_{CSP}(L) = {c\over 6} \log\( {e^{hL}-1\over h} \) + c',
  \label{eq:S_prediction}
\eeq
which is checked in Fig. \ref{fig:S_prediction} (left) for low values of $h$,
although its validity ranges far beyond that regime close to the conformal
point.

Expression \eqref{eq:S_prediction} can be expanded in the limit when
$hL$ is large enough as
\beq
  S_{CSP}(L) \approx {c\over 6} hL,
  \label{eq:volumetric}
\eeq
in agreement with numerical estimation (15) of \cite{Ramirez.rainbow.14},
\beq
  S_L \approx -0.318\; L\; \log(\alpha), 
\eeq
since $h/2=-\log(\alpha)$. Notice also that in the limit $h\to 0^+$, one
recovers the expression \eqref{eq:S_conformal}.

It is worth to compare Eq. \eqref{eq:S_prediction} with the entropy of a
thermal state at inverse temperature $\beta=1/T$ in a CFT
\cite{Calabrese.JSTAT.04}
\beq
  S_{CFT}(L)\approx \frac{c}{3} \log\( \frac{\beta}{\pi} \sinh \(
  \frac{\pi L }{\beta}   \) \) \approx \frac{\pi c L}{3\beta}, 
  \label{eq:cft_finite_t}
\eeq
where we have taken the limit $L\gg \beta$, which leads to an
extensive entropy. Comparing Eq. \eqref{eq:cft_finite_t} and
Eq. \eqref{eq:volumetric} we obtain that
\beq
  T=\frac{1}{\beta} =\frac{h}{2\pi}, 
  \label{eq:temp_again}
\eeq
in agreement with Eq. \eqref{eq:effective_temp}, which was based on the
analytic extension of the transformation employed to derive the continuum
limit, Eq. \eqref{eq:Transfx}. In other words, we can assert that the rainbow
state is similar to a thermal state with temperature given by $h$.

\subsection{R\'enyi entropies}
\label{sec:Renyi}

Let us focus our attention on R\'enyi entropies, defined as:
\beq
  S^{(n)}_A =\frac{1}{1-n} \log {\rm Tr}_A \rho^n_A,
  \label{eq:RenyiDef}
\eeq
where $A$ is a block with $\ell$ sites and $\rho_A$ is the corresponding
density matrix. The von Neumann entropy can be obtained as the limit $n \to 1$
of $S^{(n)}_A$. The expression of $S^{(n)}_A$ for the GS of the free fermion
model on an open chain of length $2L$ is given by
\beq
  S_\ell^{(n)} \simeq \frac{c}{12} \( 1 +\frac{1}{n} \) \log{\[
    \frac{4L}{\pi} \sin{ \(\frac{\pi \ell}{2L} \)} \]}
   + c'_n+ f_n \cos{(\pi \ell)} \[ \frac{8L}{\pi} \sin{\(  \frac{ \pi \ell}{2
      L}\)} \]^{-K/n},
\label{eq:RenyiClean}
\eeq
where $\ell$ is the length of the block $A$ that is located at either boundary
of the chain. The first term is the familiar CFT contribution with $c=1$,
while the second term contains the fluctuations at Fermi momentum
$k_F=\pi/2$. $K$ is the Luttinger parameter which in our case is equal to
$1$. The constants $f_n$ have also been computed analytically in reference
\cite{Fagotti.JSTAT.11}
\ba
\nonumber
  f_1 &=& -1 \\\label{fn}
  f_n &=& \frac{2}{1-n} \frac{  \Gamma \left( \frac{1}{2} +
  \frac{1}{2 n} \right)}{\Gamma \left( \frac{1}{2} + \frac{1}{2 n} \right) }
\qquad (n>1).
\ea
 
Let us consider the left-half block. According to \eqref{eq:RenyiClean}, the
R\'enyi entropy of order $n$ is given by
\beq
  S_L^{(n)} \simeq \frac{1}{12} \( 1 +\frac{1}{n} \) \log{ \( \frac{4L}{\pi} 
    \)} +   c'_n + f_n \cos(\pi L) \( \frac{8 L}{\pi} \)^{-1/n}.
  \label{eq:RenyiHalfClean}
\eeq 

\begin{figure}
  \centering
  \begin{minipage}[t]{60mm}
    \includegraphics[width=60mm]{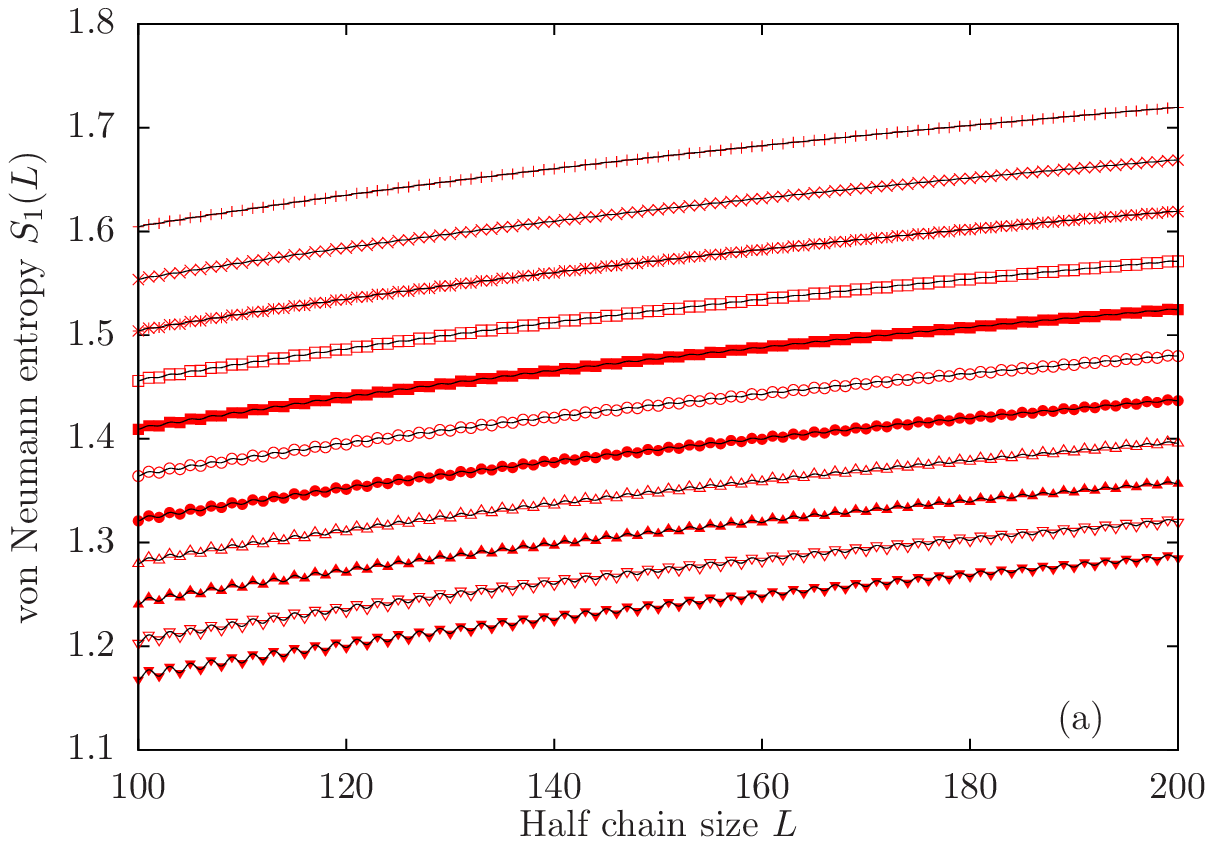}
  \end{minipage}\hspace{5mm}%
  \begin{minipage}[t]{60mm}
    \includegraphics[width=60mm]{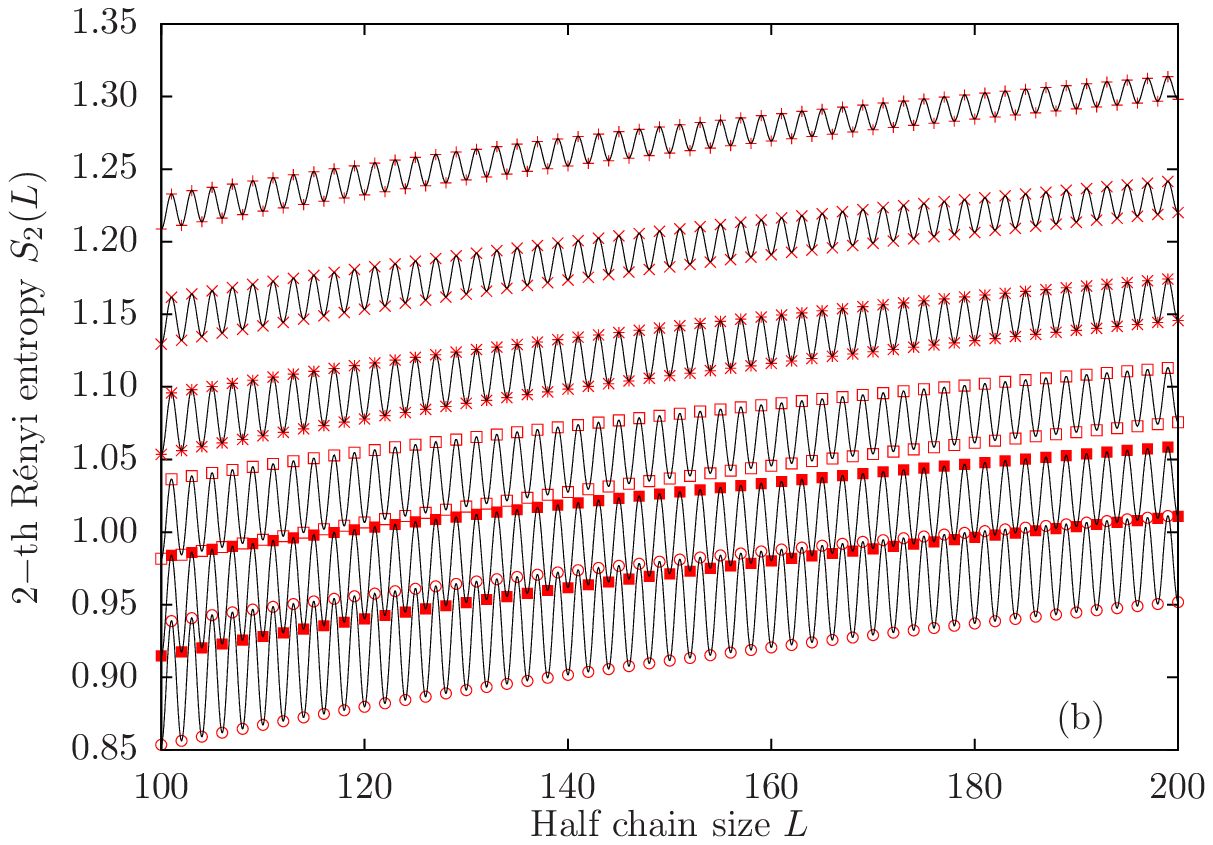}
  \end{minipage}\\\vspace{5mm}
  \begin{minipage}[t]{60mm}
    \includegraphics[width=60mm]{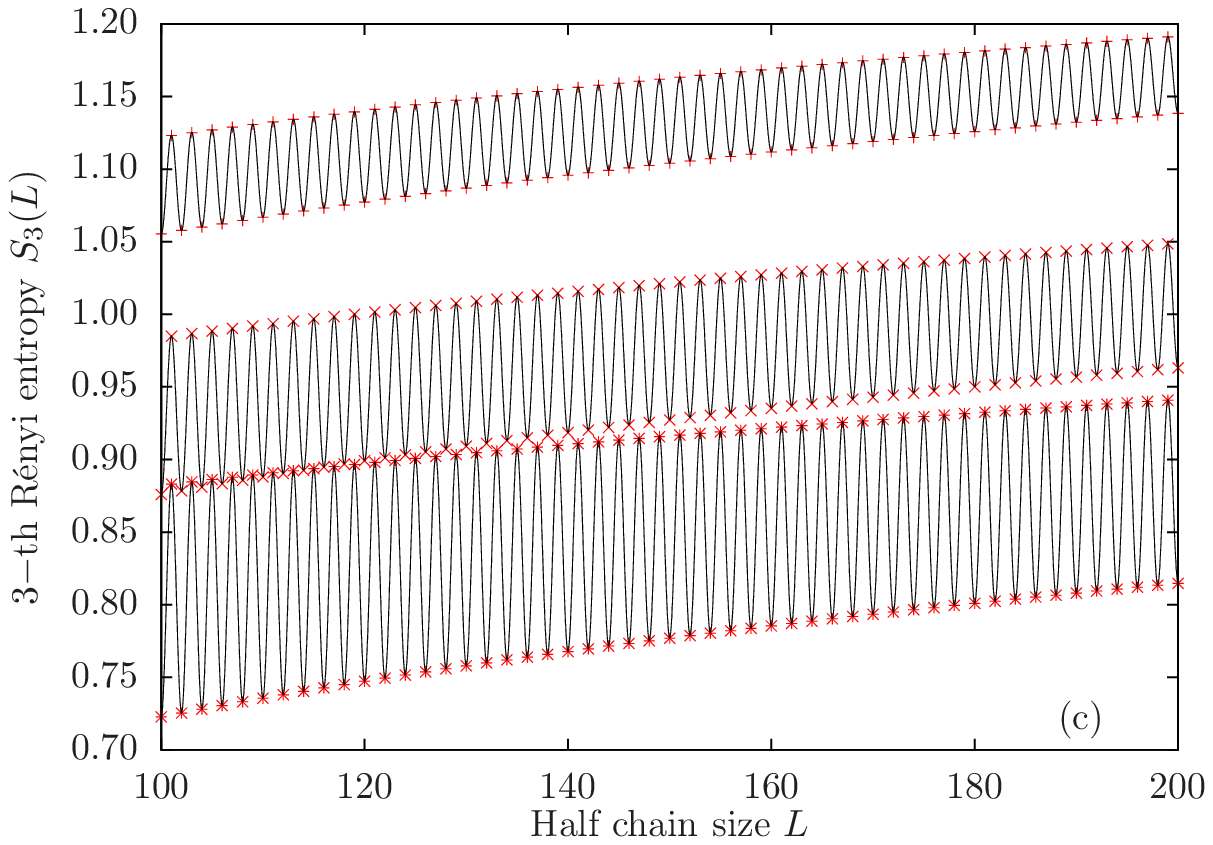}
  \end{minipage}\hspace{5mm}%
  \begin{minipage}[t]{60mm}
    \includegraphics[width=60mm]{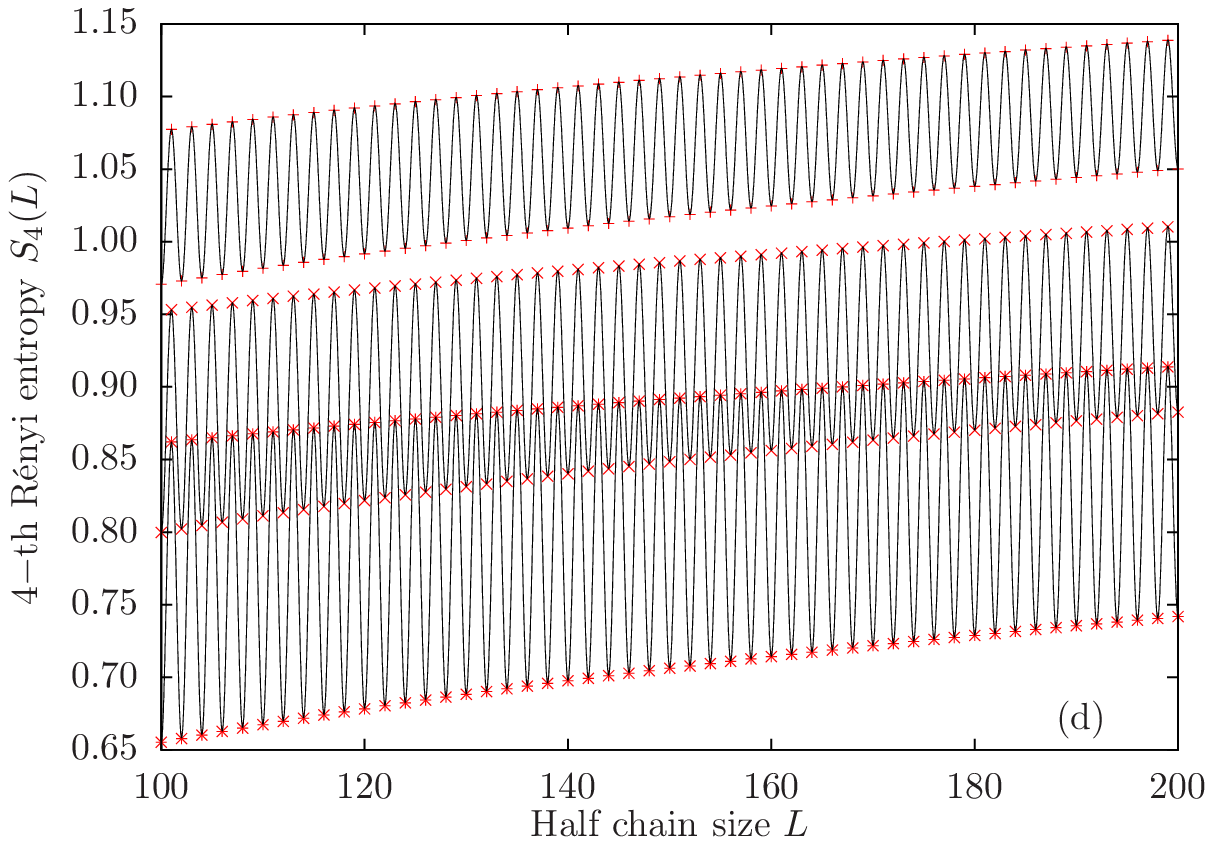}
  \end{minipage}
  \caption{R\'enyi entanglement entropy at half chain. All panels show
    $z=\{0,0.4,0.8,\cdots,4\}$ from bottom to top, and fits are made to
    expression \eqref{eq:RenyiHalf}. From top to bottom, and left to right,
    orders $n=1$ (von Neumann), $2$, $3$ and $4$ of the R\'enyi entropy.}
  \label{fig:EESn}
\end{figure}

Moving into the rainbow phase, we can give a first estimate of the R\'enyi
entropies using the SDRG prescription. According to it, they are all equal
among themselves, and equal to the von Neumann entropy
\cite{Ramirez.rxx.14}. This approximation becomes exact only in the $\alpha\to
0^+$ limit. Otherwise, we should make use of the following exact
diagonalization strategy \cite{Ramirez.rxx.14,Peschel.JPA.03}:

\begin{enumerate}
\item Obtain the occupied single-body wavefunctions $\psi^k$.
\item Compute the correlation matrix, $C_{ij}=\bra<GS|c^\dagger_i
  c_j\ket|GS>=\sum_k \bar\psi^k_i \psi^k_j$, with $i$ and $j$ both inside the
  considered block.
\item Diagonalize matrix $C$ and obtain its eigenvalues $\{\nu_p\}$.
\item The R\'enyi entropy is then given by $S^{(n)}_B = {1\over 1-n} \sum_p
  \log( \nu_p^n + (1-\nu_p)^n)$.
\end{enumerate}

The numerical computations performed as above can be compared to a natural
extension of expression \eqref{eq:RenyiHalfClean}:
\beq
  S_L^{(n)}(z) =\frac{c_n(z)}{12}\(1+\frac{1}{n}\) \log{\(\frac{4L}{\pi}\)}
  +d_n(z) 
  + f_n(z)(-1)^L\(\frac{8L}{\pi}\)^{-1/n} .
\label{eq:RenyiHalf} 
\eeq
This equation is a generalization of the Ansatz made in reference
\cite{Ramirez.rainbow.14} for the von Neumann entropy of the half-chain
$S_L^{(1)}(z)$. The comparison is performed in Fig. \ref{fig:EESn}, which
shows the R\'enyi entropies at half chain for different values of $z$ in each
panel, fitting the parameters $c_n(z)$, $d_n(z)$ and $f_n(z)$. The Luttinger
constant is kept as $K=1$. Oscillations in all cases decrease as $z$
increases, but they always increase with the R\'enyi order $n$.

\begin{figure}
  \centering
    \includegraphics[width=55mm]{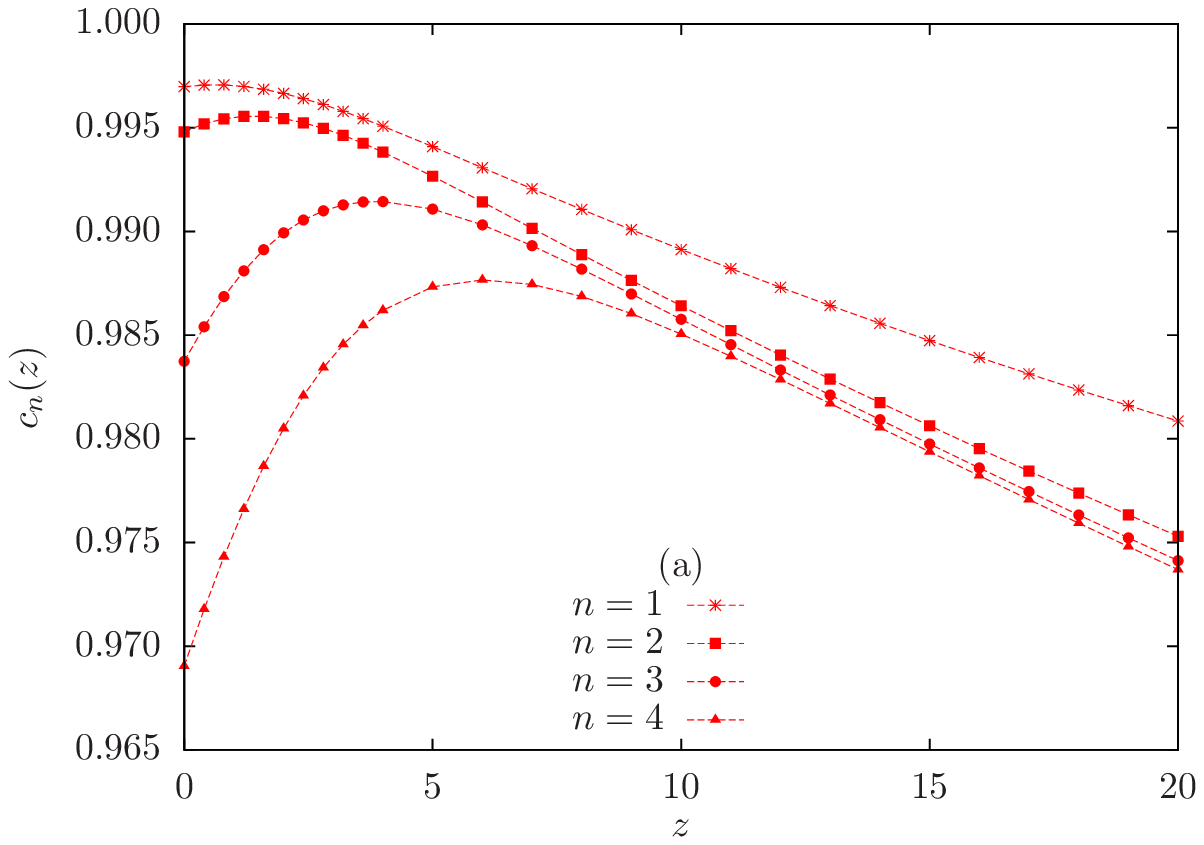}
    \includegraphics[width=55mm]{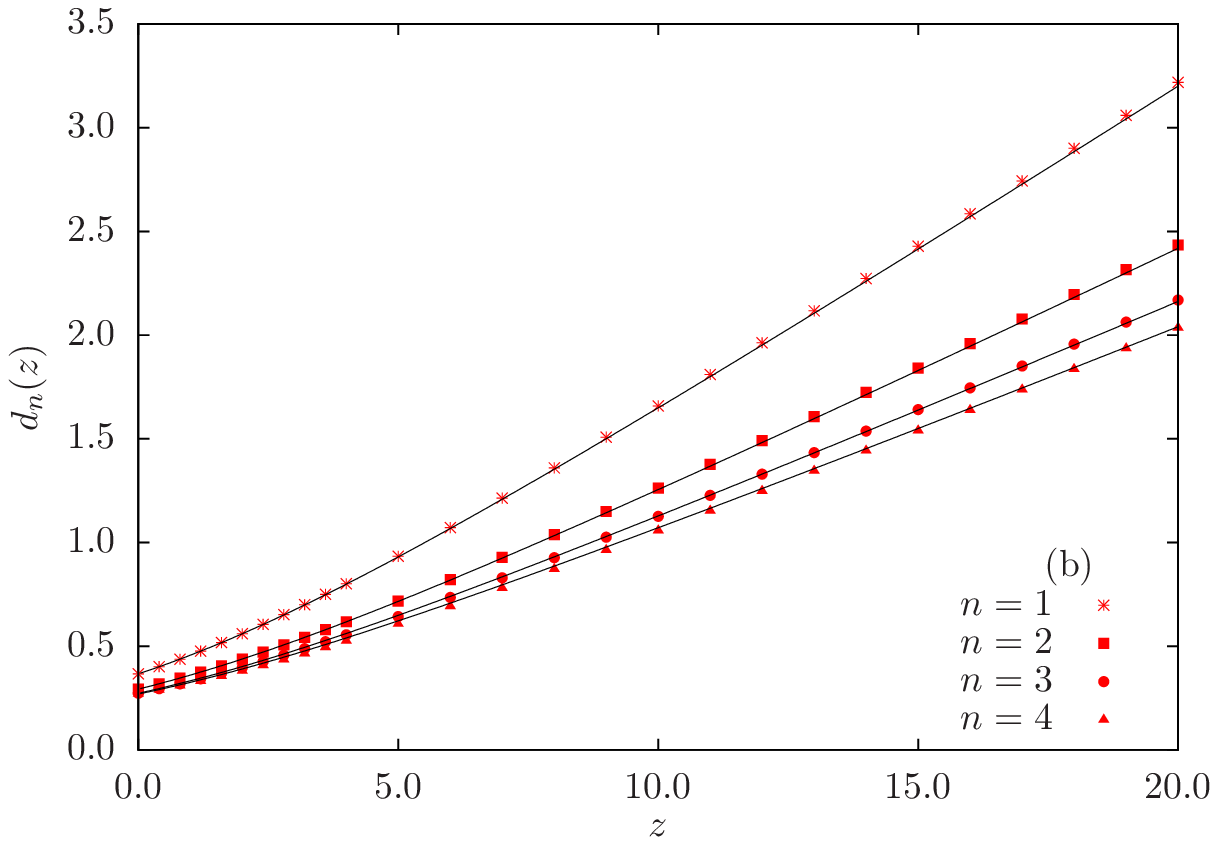}
    \includegraphics[width=55mm]{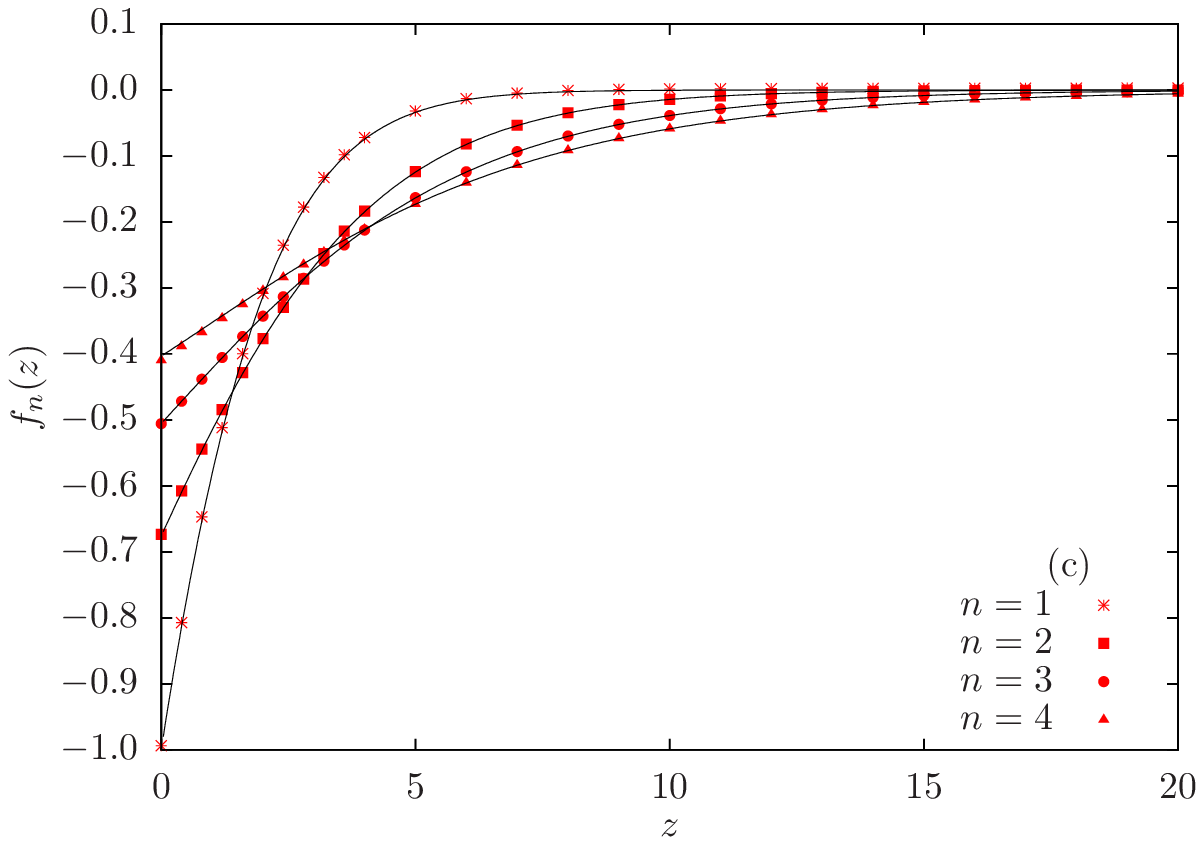}
    \caption{Fitting parameters in Eq. \eqref{eq:RenyiHalf} as a function of
      $z$: (a) $c_n(z)$, where the lines are for visual guide only; (b)
      $d_n(z)$, where the continuous line is given in Eq. \eqref{cdf} (c)
      $f_n(z)$, where the continuous line is given in Eq. \eqref{cdf}.}
  \label{fig:EEfits}
\end{figure}

The functions $c_n(z)$, $d_n(z)$ and $f_n(z)$, are shown in
Fig. \ref{fig:EEfits}. Their expression can be derived replacing $L$ by
$\tilde{L}$ in Eq. \eqref{eq:TransfL}, and writing
\beq
  \tilde{L} = \frac{e^z -1}{z} L, 
  \label{newL}
\eeq
which yields  
\beq
  c_n(z) =1, \qquad d_n(z)  = c'_n +  \frac{c}{12}\(1+\frac{1}{n}\)  \log \left(
    \frac{ e^z -1}{z} \right), \quad 
  f_n(z) = f_n \left( \frac{ e^z -1}{z} \right)^{-1/n},
  \label{cdf}
\eeq

Fig. \ref{fig:EEfits} shows the fitting coefficients for different orders of
the R\'enyi entropy, Eq. \eqref{eq:RenyiHalf}, for systems of size $10^2\leq
L\leq 10^3$ and for a range of values $0\leq z \leq 20$. Panel (a) shows the
small variation ($<4\cdot 10^{-2}$) for $c_n(z)$ in all range of $z$.  Panels
(b-c) show $d_n(z)$ and $f_n(z)$, solid lines are given by
Eq. \eqref{cdf}. Notice the perfect agreement between these expressions and
the numerical results.

\subsection{Entanglement spectrum and thermofield states}
\label{sec:EntSpectrum}

The entanglement spectrum (EE) of this model was already considered in
\cite{Ramirez.rainbow.14}. It is given by the eigenvalues of the entanglement
Hamiltonian $H_E$, that is defined in terms of the reduced density matrix
$\rho_A$ as $\rho_A = e^{-H_E}$. Let us summarize the results obtained in
\cite{Ramirez.rainbow.14} and provide an interpretation from the perspective
obtained in the previous sections. The entanglement Hamiltonian $H_E$ has the
form
\beq
H_E = \sum_{p}  \epsilon_p b^\dagger_p \, b_p + f_0, 
\label{he}
\eeq
where $b_p, b^\dagger_p$ are destruction and annihilation fermion operators,
and $\epsilon_p$ are related to the eigenvalues of the correlation matrix in
the block $A$ (for details see \cite{Ramirez.rainbow.14}). For large values of
$L$, the energies $\epsilon_p$ are given approximately by
\beq
  \epsilon_p \simeq \Delta_L \, p, 
  \quad p = \left\{
    \begin{array}{cc} 
      \pm \frac{1}{2}, \pm \frac{3}{2}, \dots, \pm \frac{L-1}{2} 
      & L: {\rm even} , \\
      0, \pm 1, \pm 2, \dots, \pm \frac{L-1}{2} 
      & L: {\rm odd}.  \\
    \end{array}. 
  \right.  
  \label{ep}
\eeq

The level spacing $\Delta_L$ in turn was found in \cite{Ramirez.rainbow.14} to
be related to the half-chain entropy $S_L$ as
\beq
  S_L \approx \frac{ \pi^2}{3 \Delta_L}, 
  \label{SL}
  \eeq 
which, using Eq. \eqref{eq:TransfL} and \eqref{eq:effective_temp}, implies
\beq
  \Delta_L \approx \frac{2 \pi^2}{h L} = \frac{\pi \beta}{L}. 
  \label{SL2}
\eeq

\begin{figure}
\epsfig{file=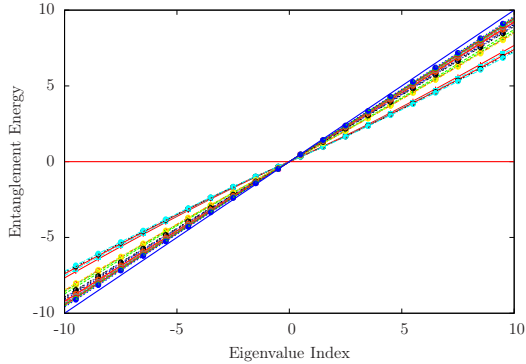,width=8cm}
\caption{\label{fig:ES} Entanglement single-body energies $\epsilon_p$ for
  different values of $L$ (60 to 160) and $z$ ($5$ to $L$ in each case),
  multiplied by $z/(2\pi^2)$. As predicted, they collapse to the diagonal
  line, following Eq. \eqref{eq:ES_collapse}, $\epsilon_p \simeq 2\pi p/z$.}
\end{figure}

Hence, the density matrix can be expressed as
\beq
  \rho_A \approx  e^{-\beta H_{CFT}}, \qquad 
  H_{CFT} = \frac{\pi}{L} \sum_{p}  p \,  b^\dagger_p \, b_p, 
  \label{SL3}
\eeq 
where $H_{CFT}$ is the CFT Hamiltonian of half of the chain. Thus, the
single-body entanglement energies should fulfill, for different values of $L$
and $z$, the following law:

\beq
  \epsilon_p \simeq \beta\, \epsilon_p^{CFT} = \( {2\pi \over h} \)\,
  \( {\pi\over L}\, p \) = {2\pi^2\over z}\, p,
  \label{eq:ES_collapse}
\eeq
which we can see confirmed in the results of Fig. \ref{fig:ES}.

We then arrive at the conclusion that the CSP state can be written as
\beq
  \ket|\psi_{CSP}> = \sum_n e^{-\beta E_n/2} \ket|n>_L \, \ket|n>_R, 
  \label{SL4}
\eeq
where $\ket|n>_R$ and $\ket|n>_L$ are orthonormal basis for the right and left
pieces of the chain whose Hamiltonians are isomorphic to $H_{CFT}$ in
Eq. \eqref{SL3}. A pure state of the form \eqref{SL4} is called a thermofield
state and has been employed in connection with black holes and the EPR=ER
conjecture \cite{Hartman.JHEP.13,Maldacena.FPhys.13}. In these studies,
$T=1/\beta$ is the temperature of the black hole that can be expressed as
$a/(2 \pi)$ where $a$ is the acceleration of a Rindler observer.  Looking at
Eq. \eqref{eq:temp_again}, we see that the constant $h$ plays that role in our
model.

Numerical evidence for other cases where $\rho\sim e^{- H_{CFT}}$ was explored
before using different scalings \cite{Lauchli.13}.

%%%%%%%%%%%%%%%%%%%%%%%%%%%%%%%%%%%%%%%%%%%%%%%%%%%%%%%%%%%%%%%%%%%%%%%%%%
\section{Two-dimensional extension}
\label{sec:rainbow2D}

A natural question is: can the 1D results be extended to 2D? In other terms,
can we find a local 2D Hamiltonian whose GS violates maximally the area law?
We shall next show that this is indeed possible in a rather simple way.

\begin{figure}
  \centering
  \epsfig{file=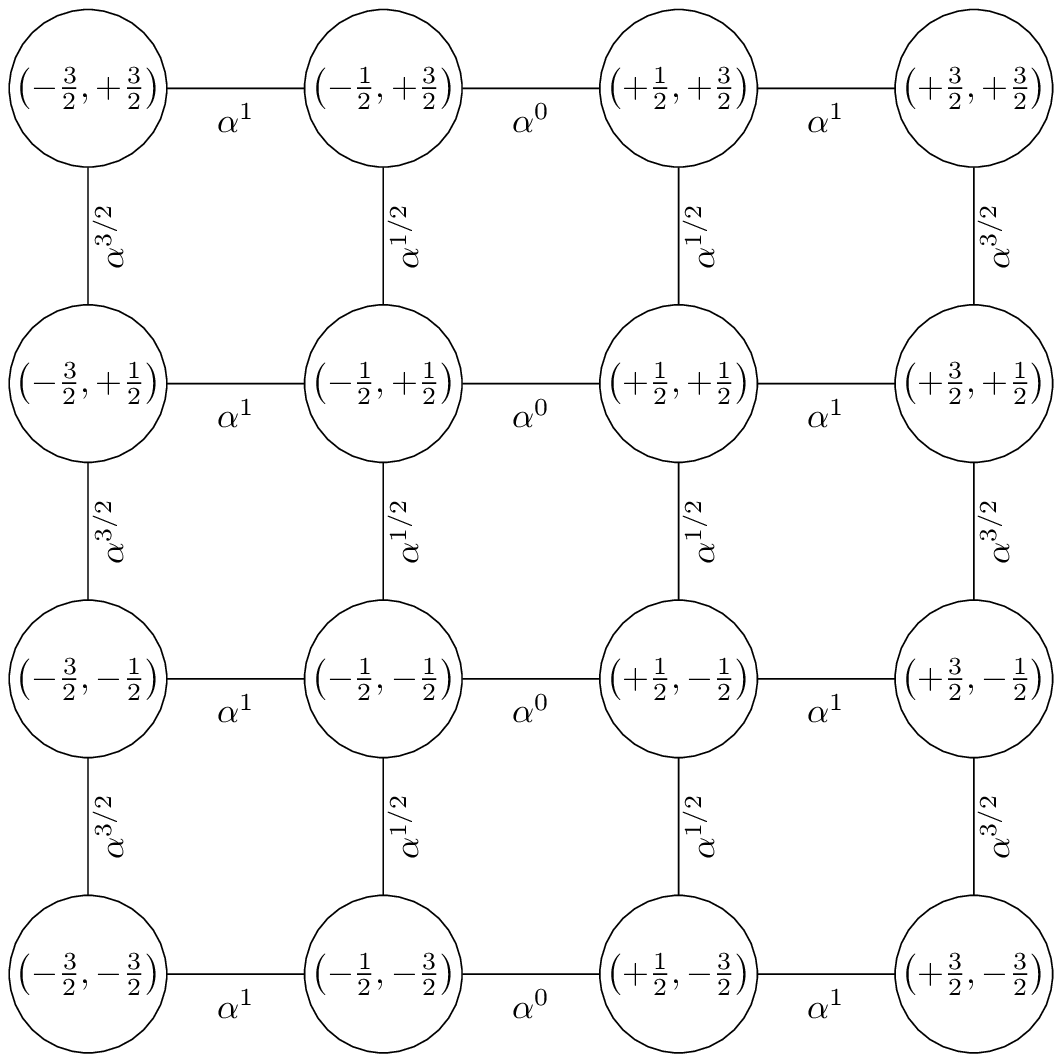,width=6cm}
  \epsfig{file=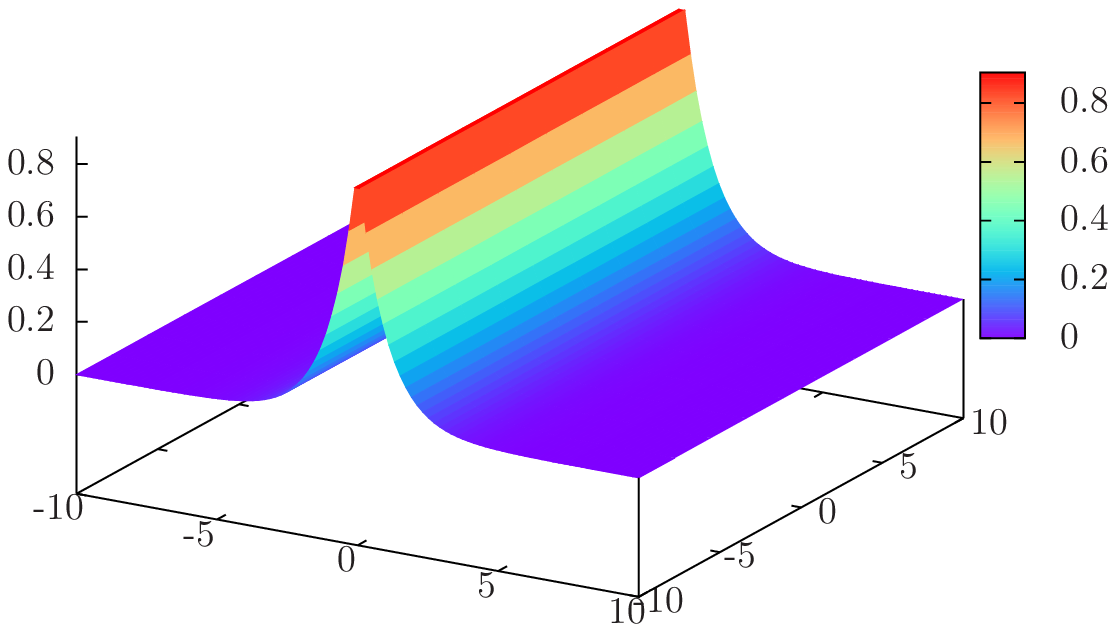,width=8cm}
  \caption{Left: a small region of the $2L\times 2L$ square lattice used to
    generate the 2D rainbow state. The nodes represent the sites, and the
    number attached to the link is the associated hopping amplitude, given by
    $\alpha^{|x|}$, with $x$ the horizontal coordinate of the middle
    point. Right: 3D view of the structure of the hopping distribution.}
  \label{fig:2dgrid}
\end{figure}

Let us consider a $2L \times 2L$ square lattice whose sites are labeled by
$X=(x,y)$ with $x,y\in \{ \pm 1/2, \pm 3/2,\cdots, \pm(L/2-1/2)\}$. We define
a hopping Hamiltonian of the form:
\beq
  \label{eq:H2D}
  H=-\sum_{\<X,X'\>} J_{X,X'} c^\dagger_{X} c_{X'}
  +\mathrm{h.c.}
\eeq
with $J_{X,X'} =F\( (X+X')/2 \)$ is only determined by the center of the
segment joining points $X$ and $X'$. In our case, we choose $F(x,y)
=\alpha^{|x|}$, to resemble the 1D analogue. Fig. \ref{fig:2dgrid} represents
graphically a small region near the center.

\begin{figure}
  \centering
  \includegraphics[width=8cm]{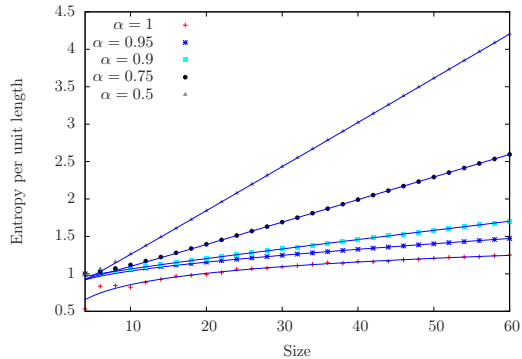}
  \caption{Entanglement entropy of the left half of the system, per unit
    length of the boundary, for different values of $\alpha$ for a 2D
    concentric singlet phase.}
  \label{fig:rainbow2d}
\end{figure}

Fig. \ref{fig:rainbow2d} shows the entropy per unit length of a block
composed of the left half of the system $s_L(\alpha)=S_L(\alpha)/L$
for different values of the coupling parameter $\alpha$. The solid
lines represent fits to an expression of the form
\beq
  s_L(\alpha) \simeq A(\alpha) L + B(\alpha) \log(L) + C(\alpha),
  \label{eq:SL2d}
\eeq
where a non-zero linear term denotes a volumetric behavior of the entanglement
entropy, and the logarithmic term is added in order to predict the correct
behavior for $\alpha\to 1^-$ \cite{Klich.06}. The fits can be seen in table
\ref{tab:Fits2d}. Notice the low values of $\chi^2$ for $\alpha<1$ and the
increase of the volume coefficient with $\alpha$.

\begin{table}
  \caption{\label{tab:Fits2d} Fitting parameters for the 2D entropy function
    in equation \eqref{eq:SL2d}}
 \begin{tabular}{@{}ccccc}
   \hline
      $\alpha$ & $A(\alpha)$ & $B(\alpha)$ & $C(\alpha)$ & $\chi^2$\\
      \hline
      1     & 0          &  0.234   &  0.29   & $3\cdot 10^{-5}$ \\
      0.95  & 0.0053     &  0.0940  &  0.77   & $10^{-9}$        \\
      0.9   & 0.0116     &  0.0330  &  0.87   & $10^{-9}$        \\
      0.75  & 0.0307     & -0.0225  &  0.85   & $10^{-9}$        \\
      0.5   & 0.0594     & -0.015   &  0.70   & $10^{-9}$        \\
      \hline
    \end{tabular}
\end{table}

%%%%%%%%%%%%%%%%%%%%%%%%%%%%%%%%%%%%%%%%%%%%%%%%%%%%%%%%%%%%%%%%%%%%

\section{Conclusions and further work}
\label{sec:conclusions}

In this work we have extended the previous studies of the concentric singlet
state, or rainbow state \cite{Vitagliano.NJP.10,Ramirez.rainbow.14}, which is
a deformation of a 1D system which exhibits a volume law for the entanglement
entropy. We focus on a free fermion realization in order to benefit from the
exact solubility. The extension presented in this article is based on the
application of field-theory methods. We show that, in the vicinity of the
conformal model, the ground state can be described by a map that is the union
of the conformal maps associated to each half of the chain. The corresponding
conformal transformations further suggest the definition of a temperature that
is proportional to the parameter controlling the decay of the hopping
parameters. We show how this deformation accounts for the change in the
dispersion relation, the single-particle wavefunctions in the vicinity of the
Fermi point, and the half-chain von Neumann and R\'enyi entropies. The
appearance of a volume law entropy is linked to the existence of an effective
temperature for the GS that is finally identified with a thermofield
state. This striking result points towards an unexpected connection with the
theory of black holes and the emergence of space-time from entanglement.
 
Finally, we show how to extend the rainbow Hamiltonian to several dimensions
in a natural way, and check that the entanglement entropy of the
two-dimensional analogue grows as the area of the block.

%%%%%%%%%%%%%%%%%%%%%%%%%%%%%%%%%%%%%%%%%%%%%%%%%%%%%%%%%%%%%%%%%%%%%

\begin{acknowledgments}
  We would like to acknowledge P. Calabrese, J.I. Cirac, E. Tonni,
  J.I. Latorre, A. L\"auchli, J. Molina, M. Ib\'a\~nez-Berganza and
  J. Simon. We acknowledge financial support from the Spanish government from
  grant FIS2012-33642, the Spanish MINECO Centro de Excelencia Severo Ochoa
  Programme under grant SEV-2012-0249 and QUITEMAD+
  S2013/ICE-2801. J.R.-L. acknowledges support from grant
  FIS2012-38866-C05-1. G.R. acknowledges support from grant FIS2009-11654.
\end{acknowledgments}

%%%%%%%%%%%%%%%%%%%%%%%%%%%%%%%%%%%%%%%%%%%%%%%%%%%%%%%%%%%%%%%%%%%%%%%

\appendix
\section{Qubistic picture}
\label{sec:qubism}

Qubism \cite{Laguna.NJP.12} is a pictographic representation for quantum
many-body states with the peculiarity that it allows for the visualization of
entanglement. In summary, an $N$ qubit wavefunction is shown on the $[0,1]^2$
square divided into $2^{N/2}\times 2^{N/2}$ cells. Each of the $2^N$
wavefunction component are depicted into one of the cells, following a
recursive pattern, in which the $i$-th qubit is associated with the $i$-th
length scale, in decreasing order.

\begin{figure}
  \centering 
\centerline{\includegraphics[width=40mm]{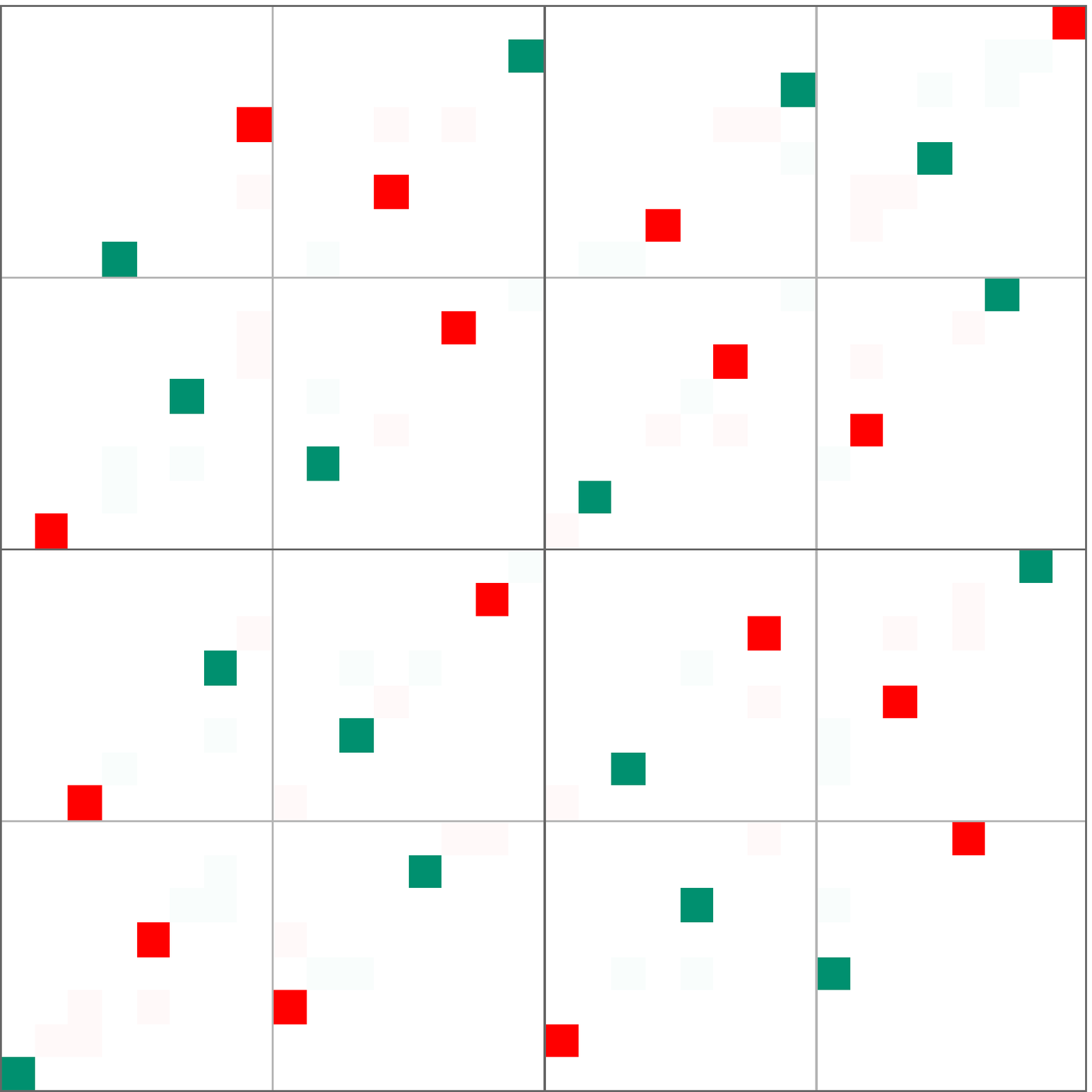}
  \includegraphics[width=40mm]{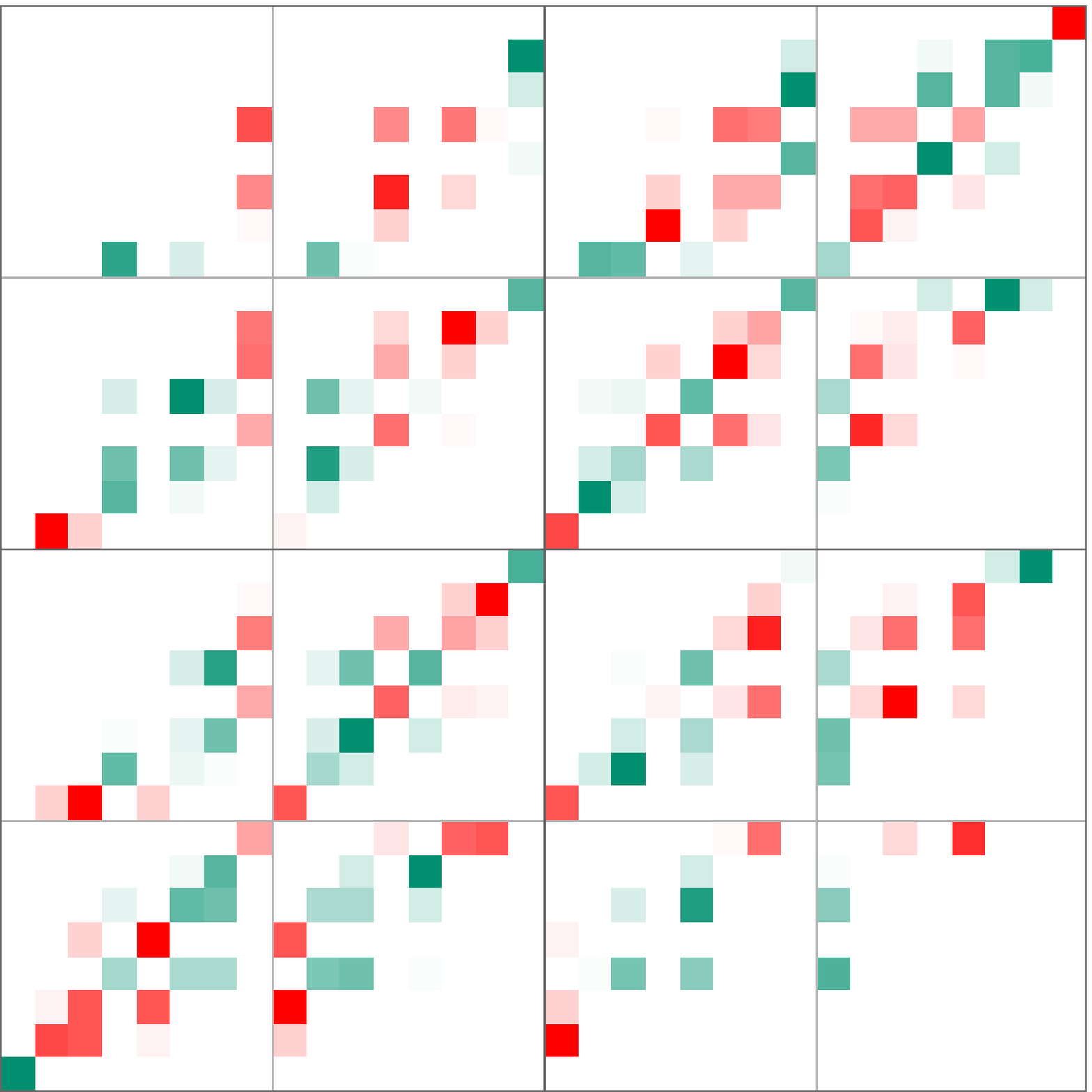}
  \includegraphics[width=40mm]{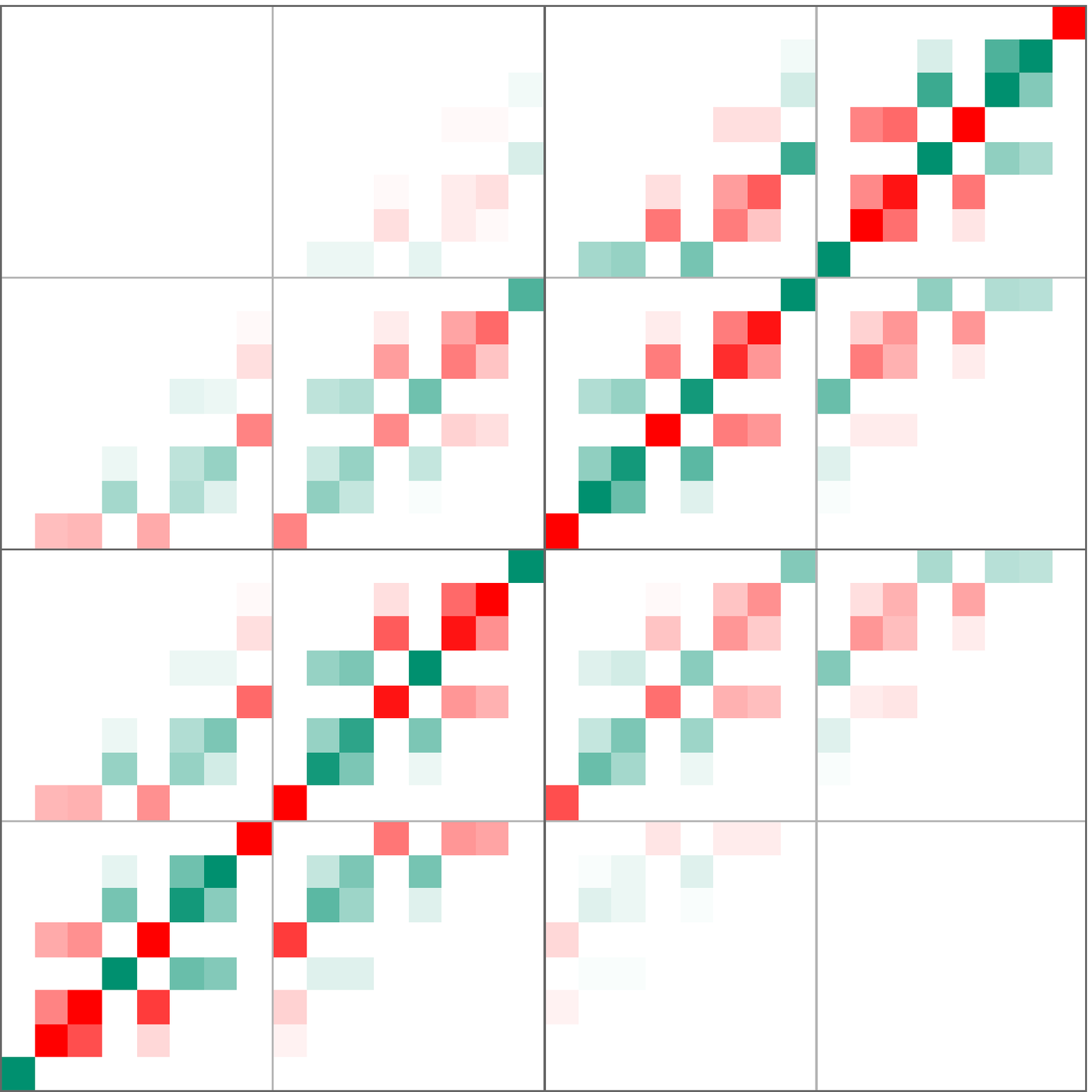}}
\centerline{\includegraphics[width=40mm]{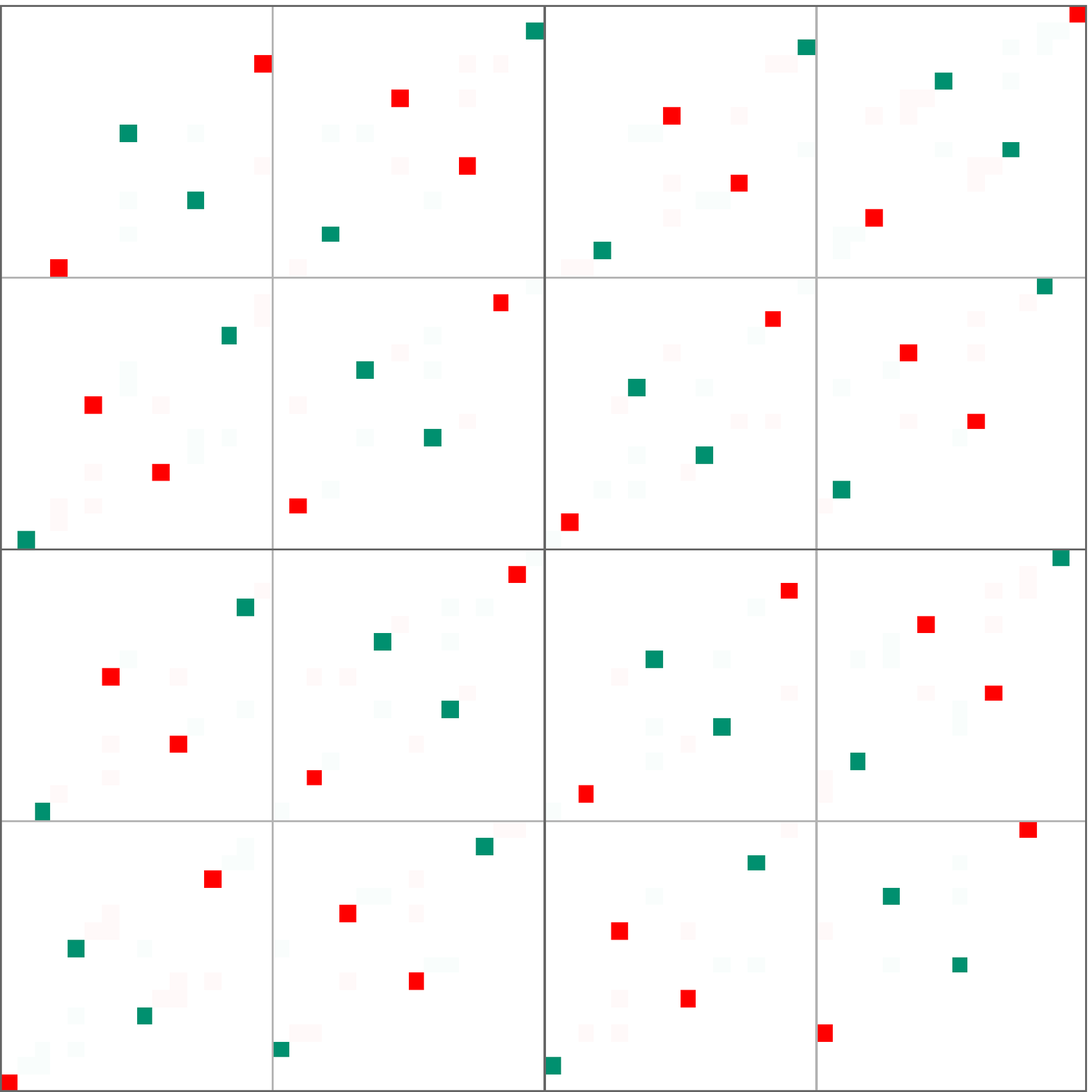}
    \includegraphics[width=40mm]{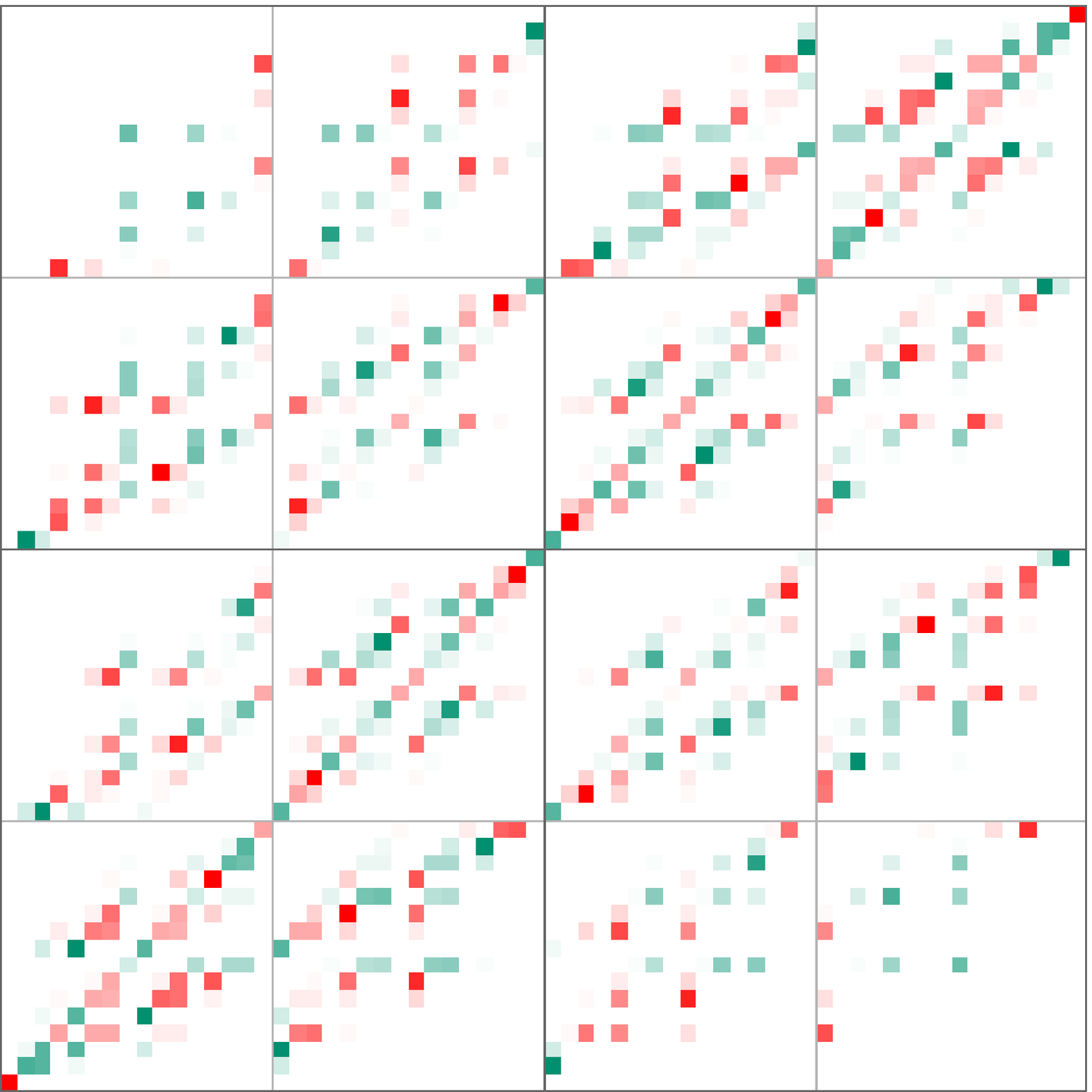}
    \includegraphics[width=40mm]{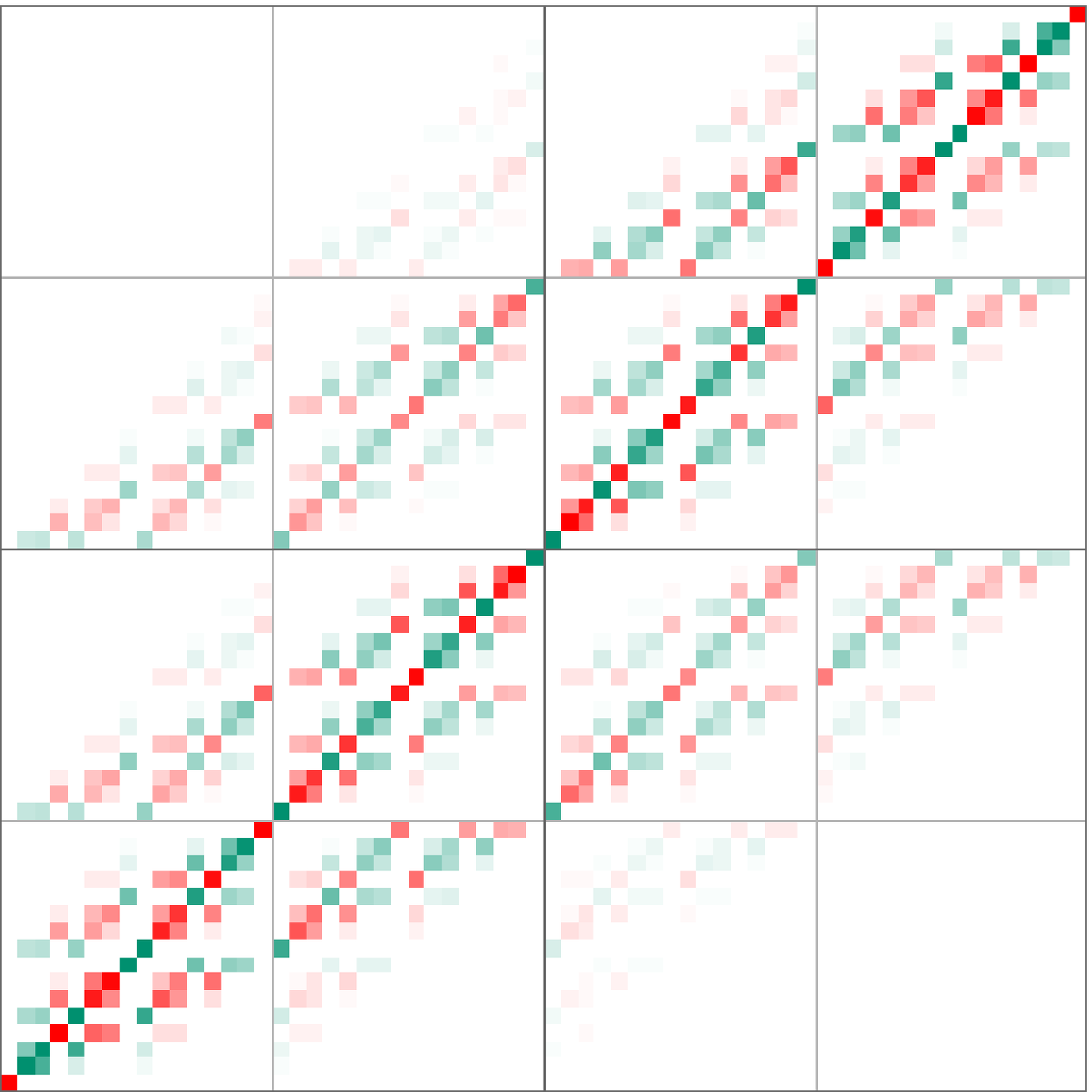}}
  \caption{Qubistic picture of the rainbow state for different sizes of the
    system $2L=10$ and $2L=12$ (upper and lower rows) and three values of
    $\alpha=0.01$, $0.3$ and $1$ (left to right columns). Color intensity
    denotes the wavefunction amplitude and hue denotes phase: red is positive
    and green negative.}
  \label{fig:qubism}
\end{figure}

Fig. \ref{fig:qubism} represents the qubistic plots of the rainbow ground
state for two different sizes, $2L=10$ and $2L=12$, and three values of
$\alpha=0.01$, $0.3$ and $1$. Therefore, the rightmost panels correspond to
the ground state of the free fermion model, and the leftmost panels represent
the rainbow states. Notice that the representation is formed only by a finite
and small set of points.

Entanglement between the first pair of qubits and the rest can be visualized
in the following way \cite{Laguna.NJP.12}. Break the full square into $2\times
2$ square of half-size. Count the number of different (strictly, linearly
independent) images among the small squares. That number is an upper bound for
the Schmidt rank, which is a measure of entanglement. The same procedure can
continue, for the block composed of the first four qubits, if we decompose the
original square into a $4\times 4$ grid. In our case, notice that the dots in
each of the small squares form a similar but different pattern. In fact, the
number of different (independent) images coincides with the number of squares,
$4$ for the first two qubits, $16$ for the first four, etc. This shows that
the Schmidt rank grows as $2^\ell$, i.e., entanglement is maximal.

%%%%%%%%%%%%%%%%%%%%%%%%%%%%%%%%%%%%%%%%%%%%%%%%%%%%%%%%%%%%%%%%%%%%%


\begin{thebibliography}{10}
\bibitem{footnote1} In order to obtain further insight into the reason for
  this transformation, consider a one-dimensional chain of $L=4$ sites,
  numbered $1$ to $4$, where two singlet bonds are established, between sites
  $1$ and $4$, and between sites $2$ and $3$. When transformed, via the
  Jordan-Wigner transformation, into a fermionic state, it becomes
  $(c^\dagger_1 + c^\dagger_4) (c^\dagger_2 - c^\dagger_3) \ket|0>$.
\bibitem{footnote3} Due to the different normalization conventions, the
  parameter $z=- L \log \alpha$ used in \cite{Ramirez.rainbow.14} is equal to
  $z/2$ used in this paper.
\end{thebibliography}
\end{document}